\documentclass[journal=jacsat,manuscript=article]{achemso}
\usepackage[version=3]{mhchem}
\usepackage[utf8]{inputenc}
\usepackage[T1]{fontenc}

\author{Peiyuan Gao}
\affiliation{Pacific Northwest National Laboratory, Richland 99352, USA}
\author{Xiu Yang}
\email{xiy518@lehigh.edu}
\affiliation{Department of Industrial and Systems Engineering, Lehigh University, Bethlehem, PA 18015, USA}
\author{Yu-Hang Tang}
\affiliation{Lawrence Berkeley National Laboratory, Berkeley, California 94720, USA}
\author{Muqing Zheng}
\affiliation{Department of Industrial and Systems Engineering, Lehigh University, Bethlehem, PA 18015, USA}
\author{Amity Anderson}
\affiliation{Pacific Northwest National Laboratory, Richland 99352, USA}
\author{Vijayakumar Murugesan}
\affiliation{Pacific Northwest National Laboratory, Richland 99352, USA}
\email{vijay@pnnl.gov}
\author{Aaron Hollas}
\affiliation{Pacific Northwest National Laboratory, Richland 99352, USA}
\author{Wei Wang}
\affiliation{Pacific Northwest National Laboratory, Richland 99352, USA}
\email{wei.wang@pnnl.gov}

\title[An \textsf{achemso} demo]
{Graphical Gaussian Process Regression Model for Aqueous Solvation Free Energy Prediction of Organic Molecules in Redox Flow Battery}
\usepackage{amsmath}
\usepackage{amsfonts}
\usepackage{amssymb}
\usepackage{enumitem}
\usepackage{mathtools}
\usepackage{bm}
\usepackage{color}
\usepackage{subfigure}
\usepackage{hyperref}

\newcommand{\tensor}[1]{\boldsymbol{#1}}
\newcommand{\mexp}[1]{\mathrm{E}\left\{#1\right\}}
\newcommand{\trans}{\top}
\newcommand{\cov}{\text{Cov}}

\newcommand{\defeq}{\vcentcolon=}

\makeatletter
    \newcommand{\smallotimes}{\mathbin{\mathpalette\make@small\otimes}}
    \newcommand{\make@small}[2]{%
    \vcenter{\hbox{%
        \scalebox{0.6}{$\m@th#1#2$}%
    }}%
    }
    \let\sv@thm\@thm
    \def\@thm{\vspace{0.75em}\let\indent\relax\sv@thm}
\makeatother

\begin{document}
\begin{abstract}
The solvation free energy of organic molecules is a critical parameter in determining emergent properties such as solubility, liquid-phase equilibrium constants, and pKa and redox potentials in an organic redox flow battery. In this work, we present a machine learning (ML) model that can learn and predict the aqueous solvation free energy of an organic molecule using Gaussian process regression method based on a new molecular graph kernel. To investigate the performance of the ML model on electrostatic interaction, the nonpolar interaction contribution of solvent and the conformational entropy of solute in solvation free energy, three data sets with implicit or explicit water solvent models, and contribution of conformational entropy of solute are tested. We demonstrate that 
%the uncertainty introduced by explicit water solvent and conformational change of solute can be handled by the ML model. 
our ML model can predict the solvation free energy of molecules at chemical accuracy with a mean absolute error of less than 1 kcal/mol for subsets of the QM9 dataset and the Freesolv database. 
To solve the general data scarcity problem for a graph-based ML model, we propose 
%an active learning(?) 
a dimension reduction algorithm based on the distance between molecular graphs, which can be used to examine the diversity of the molecular data set. 
It provides a promising way to build a minimum training set to improve prediction for certain test sets where the space of molecular structures is predetermined.
\end{abstract}

\flushbottom
\maketitle
\thispagestyle{empty}
\section*{Introduction}

%{\color{blue} Need to modify the intro by Vijay, Wei, Aaron}

Redox flow batteries (RFBs), particularly the aqueous organic RFBs (ORFBs), have gained significant interest for grid scale energy storage due to their inherent safety, flexible design, modular scale-up, and potential low cost. Critical functionalities of ORFBs such as energy density, cycling stability, and rate capability are largely impacted by the properties of the active organic species.\cite{Kwabi2020cr,Narayan2019COE} For example, the solubility of the active organic molecule dictates the energy density of an organic RFB. Therefore, the search for highly soluble (\textgreater 1M) and chemically stable redox active organic materials has recently become a critical research endeavor.\cite{GentilCOE2020} The solubility, as well as the reactivity, viscosity, and redox potential of the active organic molecules depend on intricate interactions between the solute and solvent molecules, for which the free energy of solvation is often a critical parameter.\cite{Sm2012jctc,Skyner2015pccp} Evidently, solvation free energy has often been identified as a critical descriptor in quantitative structure-property/activity relationships (QSPR/QSAR) analysis. Yet there have been comparatively few experimental values (\textless 2000) reported despite the millions of organic molecules synthesized to date. Density functional theory (DFT) and molecular dynamics (MD) simulation methods have been widely utilized for determining this prominent chemical descriptor.\cite{GP2009JPCB,TG1996jpc,DRM2017jced,Luukkonen2020jcim,Subramanian2020jcim,JD2019nc,Voityuk2020pccp} With recent advancements in implicit solvation models\cite{Cossi2003jcc,Tomasi2005cr,Lin2002ie,Klamt1995jpc} and operating functionals, the DFT and MD methodologies\cite{SD2012JCTC,KS2020JCTC,Roos2019jctc,Fan2020jcamd} provide a reliable estimate of solvation free energy with the mean-absolute-error approaching the chemical accuracy level of 1 kcal/mol. However, approximations are often used to lower computational time at the cost of accuracy.\cite{Fornari2020wcms,Sl2018sci} Furthermore, large-scale calculation of solvation free energy with high precision method through DFT and MD is computationally intractable. In view of this challenge, an artificial intelligence (AI) based prediction is needed because their computational strategies automatically improve through experience.\cite{Fornari2020wcms,Sl2018sci} Machine learning (ML) methods are capable to predict a very broad range of properties. Recently, neural network model (NN) has received new attention for predicting solvation free energy prediction.\cite{LY2015NATURE,AS2020CCE,yang2020pccp,Zubatyuk2019sa} Some of these architectures operate over fixed molecular fingerprints common akin to traditional QSPR models.\cite{Hutchinson2019jcim,Riniker2017jcim,MJ2015JCIM} However, due to the incomplete physical understanding of the structure of molecule and emergent properties, the features provided by domain experts may not include all critical design parameters in the material design. The graphical approach is a powerful tool to complement the domain experts knowledge because many features selected by domain experts are based on the computations which use the molecular structures.\cite{CC2017jcim,KY2020jc,Mahe2005jcim,Mosbach2020jcim,Na2020pccp,sz2021cs} Moreover, as molecules have arbitrary chemical composition and highly variable connectivity, useful information is difficult to be extracted from a molecule into a fixed dimensional representation. Thus, incorporating graphical approach can add important features that could be inadvertedly neglected by domain experts when designing an ML model. Naturally, a molecular structure can be represented by an undirected labeled graph that encodes both structural and functional information. The graph contains an initial feature vector and a neighbor list for each atom. The feature vector summarizes the atom's local chemical environment, including atom-types, hybridization types, and valence structures. Neighbor lists represent connectivity of the whole molecule.  
Another key question for molecular properties prediction using ML methods is lack of data, namely the data sparsity. Molecular properties data sets are different from the data sets in other applications as image recognition or natural language processing. Usually, the size of molecular properties data set that can be found is much smaller than those available for the aforementioned conventional machine learning tasks, as accurate results for molecular properties typically requires specialized instruments and measurements. Therefore, the measurement cost of a small data set is rather expensive and time-consuming. Even for some molecular properties which can be obtained by computer simulation, e.g., solvation free energy in explicit solvent, the calculations are also not cost-effective. So the amount of training data remains a challenge in the property prediction of molecules. 

Gaussian process (GP) is one of the most well studied stochastic processes in probability and statistics. Given the flexible form of data representation, GP is a powerful tool for classification and regression, and it is widely used in probabilistic scientific computing, engineering design, geostatistics, data assimilation, machine learning, etc.\cite{Hu2020joule,lei2018mssp,li2020jcp} In particular, given a data set comprising input/output pairs of locations and quantity of interest (QoI), GP regression (GPR), also known as Kriging, can provide a prediction along with a mean squared error (MSE) estimate of the QoI at any location. Alternatively, from the Bayesian perspective, GPR identifies a Gaussian random variable at any location with posterior mean (corresponding to the prediction) and variance (corresponding to the MSE).
In other words, a GP model not only provides point predictions in the form of posterior means but also estimates the uncertainty of the prediction using posterior variances.
Generally speaking, the larger the given data set size is, the closer the GPR’s posterior mean is to the ground truth and the smaller the posterior variance is. While for small data set, the performance of GPR model is also good compared with deep neural network which typically requires a large training set.\cite{Aditya2018jcp} Therefore, GP method is a good candidate for the machine leaning works when large data sets are difficult to be obtained.  

In this work, we propose a machine learning model to predict the solvation free energy of organic molecules in water. We implement a graphical-kernel-based GP method~\cite{kashima2003marginalized, tang2019prediction} to construct surrogate models for solvation free energy prediction. In contrast to previous studies\cite{CC2017jcim,KY2020jc,Mahe2005jcim,Mosbach2020jcim,Na2020pccp,sz2021cs}, a weighted and labeled graph with labels on both nodes and edges in this work is used to give a more accurate representation for the inner structure of a molecule.
Furthermore, to investigate the capability of our machine learning model on different components of solvation free energy in thermodynamics as electrostatic interaction energy, the nonpolar interaction contribution of solvent and the contribution of conformational entropy of solute, we build and test three solvation free energy data sets, namely our own Pacific Northewest National Laboratory (PNNL) organic molecule data set, the QM9 data set, and the Freesolv data set. The solvation energy data in the three data sets include either the conformational entropy contribution or the effect of explicit solvent, or both of them. Our results are benchmarked against the three data sets. We demonstrate that our ML model can predict the solvation free energy of molecules at chemical accuracy (\textless 1 kcal/mol) and 1000-10000 times faster than DFT/MD methods. Additionally, we try to elucidate the relationship between the molecular graph and molecular property using the model reduction method and provide a possible way on how to build a minimum training set to better predict the corresponding molecular property with ML model.

\section*{Method}

\subsection*{GPR framework}
\label{subsec:gpr}
We present a brief review of the GPR method adopted from Reference ~\cite{abrahamsen1997review, forrester2008engineering}.
We denote the observation locations as $\bm X = \{\bm x^{(i)}\}_{i=1}^N$
($\bm x^{(i)}\in D, D\subseteq\mathbb{R}^d$) and the 
observed values of the QoI at these locations as 
$\bm y=(y^{(1)}, y^{(2)},\dotsc, y^{(N)})^\trans$
($y^{(i)}\in\mathbb{R}$). For simplicity, we assume that $y^{(i)}$ are scalars. 
%We aim to predict $y$ at any new location $\bm x^*\in D$. 
%The GPR method assumes that the observation vector $\bm y$ is a realization of the following $N$-dimensional random vector that satisfies multivariate Gaussian distribution:
%-------------------------------------------------------------------------------
%\[\bm Y = \left(Y(\bm x^{(1)}), Y(\bm x^{(2)}), \dotsc, Y(\bm x^{(N)})\right)^\trans,\]
%-------------------------------------------------------------------------------
%where $Y(\bm x^{(i)})$ is the concise notation of $Y(\bm x^{(i)}; \omega)$,
%and $Y(\bm x^{(i)}; \omega)$ is a Gaussian random variable defined on a 
%probability space $(\Omega, \mathcal{F}, P)$ with $\omega\in\Omega$.
The GPR method aims to identify a GP $Y(\bm x,\omega):D\times\Omega\rightarrow\mathbb{R}$ based on the input/output data set $\{(\bm x^{(i)}, y^{(i)})\}_{i=1}^N$, where $\Omega$ is the sample space of a probability triple.
Here, $\bm x$ can be considered as parameters for this GP, such that 
$Y(\bm x,\cdot):\Omega\rightarrow \mathbb{R}$ is a Gaussian random 
variable for any $\bm x$ in the set $D$. A GP $Y(\bm x, \omega)$ is usually denoted as
%-------------------------------------------------------------------------------
\begin{equation}
  \label{eq:gp0}
Y(\bm x) \sim \mathcal{GP}\left(\mu(\bm x), k(\bm x, \bm x')\right),
\end{equation}
%-------------------------------------------------------------------------------
where $\omega$ is not explicitly listed for brevity, $\mu(\cdot):D\rightarrow\mathbb{R}$ and 
$k(\cdot,\cdot):D\times D\rightarrow\mathbb{R}$ are the mean 
and covariance functions (also called \emph{kernel} function), respectively:
%-------------------------------------------------------------------------------
  \begin{align}
    \mu(\bm x) & = \mexp{Y(\bm x)},\\ 
    k(\bm x,\bm x') & = \cov\left\{Y(\bm x), Y(\bm x')\right\}
                      = \mexp{(Y(\bm x)-\mu(\bm x))(Y(\bm x')-\mu(\bm x'))}.
  \end{align}
%-------------------------------------------------------------------------------
The variance of $Y(\bm x)$ is $k(\bm x, \bm x)$, and its standard deviation is
$\sigma(\bm x)=\sqrt{k(\bm x,\bm x)}$. 
The covariance matrix, denoted as $\tensor C$, is 
defined as $C_{ij}=k(\bm x^{(i)}, \bm x^{(j)})$.
%Functions $\mu(\bm x)$ and $k(\bm x,\bm x')$ are obtained by identifying their
%hyperparameters via maximizing the log marginal 
%likelihood~\cite{williams2006gaussian}:
%%-------------------------------------------------------------------------------
%\begin{equation}
%  \label{eq:lml}
%  \ln L=-\dfrac{1}{2}(\bm y-\bm\mu)^\trans \tensor C^{-1} (\bm
%  y-\bm\mu)-\dfrac{1}{2}\ln |\tensor C|-\dfrac{N}{2}\ln 2\pi,
%\end{equation}
%%-------------------------------------------------------------------------------
%where $\bm\mu=(\mu(\bm x^{(1)}),\dotsc,\mu(\bm x^{(N)}))^\trans$ and $\vert\tensor C\vert$ is the determinant of matrix $\tensor C$. 
For any $\bm x^*\in D$, the GPR prediction and variance are
%posterior mean and variance are posterior distribution 
%$Y(\bm x^*, \cdot)\sim\mathcal{N}(\hat y(\bm x^*), \hat s^2(\bm x^*))$ for , where
%-------------------------------------------------------------------------------
  \begin{align}
\label{eq:krig}
\hat y(\bm x^*) & = \mu(\bm x^*) + 
  \bm c(\bm x^*)^\trans\tensor{C}^{-1}(\bm y-\bm\mu),  \\
  \hat s^2(\bm x^*) & = 
   \sigma^2(\bm x^*)-\bm c(\bm x^*)^\trans \tensor{C}^{-1}\bm c(\bm x^*),
 \end{align}
%-------------------------------------------------------------------------------
where $\bm c(\bm x^*)$ is a vector of covariance:
$(\bm c(\bm x^*))_i=k(\bm x^{(i)},\bm x^*)$.
%In practice, it is common to use $\hat y(\bm x^*)$ as the prediction, and
Here $\hat s^2(\bm x^*)$ is also called the mean squared error (MSE) of the prediction
because $\hat s^2(\bm x^*)=\mexp{(\hat y(\bm x^*)-Y(\bm x^*))^2}$
\cite{forrester2008engineering}. 
Consequently, $\hat s(\bm x^*)$
%the posterior standard deviation, 
is called the root mean squared error (RMSE). 

%In the widely used ordinary Kriging method, a stationary GP is assumed
%\cite{kitanidis1997introduction}. Specifically, $\mu$ is set as a constant 
%$\mu(\bm x)\equiv\mu$, and $k(\bm x, \bm x')=k(\bm\tau)$, where 
%$\bm\tau=\bm x-\bm x'$. Consequently, 
%$\sigma^2(\bm x)=k(\bm x,\bm x)=k(\bm 0)=\sigma^2$ is a constant.
In practice, it is common to assume that $\mu(\bm x)$ is a constant function,
i.e., $\mu(\bm x)\equiv \mu$. Also, the most widely used kernels in scientific 
computing are the Mat\'{e}rn functions, especially its two special cases, i.e.,
exponential and squared-exponential (Gaussian) kernels. For example, the Gaussian 
kernel can be written as
$k(\bm\tau)=\sigma^2\exp\left(-\frac{1}{2}\Vert \bm x-\bm x'\Vert^2_w\right)$,
where the weighted norm is defined as 
$\displaystyle\Vert \bm x-\bm x'\Vert^2_w=\sum_{i=1}^d \left(\dfrac{x_i-x'_i}{l_i}\right)^2$. 
Here, $l_i$ ($i=1,\dotsc, d$), the correlation lengths in
the $i$ direction, are constants. More details are provided in the support
material.

\subsection*{Graph kernel}
Using a graph kernel, the physical location $\bm x$ in the aforementioned conventional GP can take the form of a graph. In this work, we use each graph to represent a molecule. Therefore, each $\bm x$ can be considered as a molecule. We use the graph kernel to define notation of the inner product between molecules and use it as the GP kernel $k(\bm x, \bm x')$. Following the notation in graph theory, we slightly modify the notation and use $G$ to replace $\bm x$ in the GPR method. The practice of using labeled graphs, with the exemplary
ball-and-stick model, to represent molecules gained popularity well before the era of machine learning.\cite{Tsuji2018cr,Gar2008cr} In this work,
we represent a molecule of $n$ atoms as an undirected graph 
$G=\{V=\{v_i\}, E =\{e_{ij}\}, i,j\in\{1,\cdots,n\}$, where each atom $i$ is represented by a vertices $v_i$ that are labeled by
chemical elements, charge, hybridization state, conjugacy, aromaticity, and hydrogen count.\cite{WH2017arxiv} Each edge $e_{ij}\in \mathbb{R}$ between vertices $i$ and j represents the bond between between the atoms and is labeled by bond order, aromaticity, conjugacy, and ring membership.  Its weight $w_{ij}$ is set by a spatial adjacency rule $\mathcal{A}(\bm r_i,\bm r_j)$, which will be introduced later. Thus, the adjacency matrix $\tensor A$ of a molecular graph is given as $A_{ij} = \mathcal{A}(\bm r_i,\bm r_j)$. Note that the edges are often supersets of the collection of covalent bonds in a molecule.

\begin{figure}[t]
    \centering
    \includegraphics[width=0.7\textwidth]{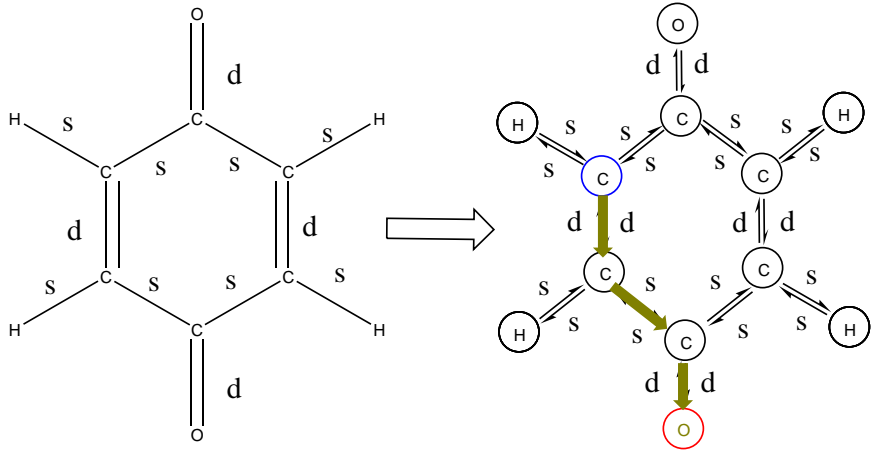}
    \caption{Demo of random walk on 1,4-benzoquinone molecule}
    \label{fig:random_walk}
\end{figure}

To implement the graph in a GP, we use the marginalized graph kernel $K(G, G')$~\cite{kashima2003marginalized}, which defines an inner product between two graphs, i.e., in our case, two molecules. The main idea is to perform random walks simultaneously on two given graphs and then calculate the expectation of the ``similarity'' between all pairs of the paths in such random walks. Specifically, each path, denoted as $\bm h$ on a graph, is the route from one atom to a certain one via chemical bonds in a molecule, and an inner product between the paths can be defined recursively using an element-wise inner products formula. Each $\bm h$ is a sequence consisting of vertices and edges:
\[v_{h_1}e_{h_1h_2}v_{h_2}e_{h_2h_3}v_{h_3}\cdots,\]
where $v_{h_k}$ is the $k$th atom traversed by this path, and $e_{h_{k-1}h_k}$ is the chemical bond connection between the $(k-1)$th and the $k$th atoms in this path.
Figure~\ref{fig:random_walk} shows an example of path between two nodes. 

The expectation of the path similarity in the simultaneous random walk is given as
\begin{equation}
\label{eq:distance}
\begin{aligned}
    K(G, G')= & \sum_{\ell=1}^{\infty}\sum_{\bm h}\sum_{\bm h} 
     \left(p_s(h_1)\prod_{i=2}^{\ell} p_t(h_i|h_{i-1})p_q(h_{\ell}) \right)\times 
     \left(p'_s(h_1')\prod_{i=2}^{\ell} p'_t(h'_j|h'_{j-1})p'_q(h'_l)\right) \\
    & \times K_v(v_{h_1}, v'_{h'_1})\prod_{k=2}^{\ell} K_v(v_{h_k}, v'_{h'_k}) K_e(e_{h_{k-1} h_k}, e'_{h'_{k-1}h'_k}).
    \end{aligned}
\end{equation}
Here, $\ell$ is the length of the path, $\bm h$ and $\bm h'$ are paths on the graphs represented by length-$l$ vectors of vertex labels, $_s(\cdot)$ is the starting probability of the random walk on each vertex, $p_q(\cdot)$ is the stopping probability of the random walk on each vertex at any given step, $p_t(\cdot | \cdot)$ is the transition probability between a pair of vertices, $K_v(\cdot,\cdot)$ is a microkernel that computes the similarity between two vertices (i.e., atoms), and $K_e(\cdot,\cdot)$ is another microkernel that computes the similarity between pairs of edges (i.e., bonds). 

Following the setup in~\cite{tang2019prediction}, we set the vertex elementary
kernel as
\begin{equation}
  \label{eq:v_kernel}
  K_v(v, v') = \begin{cases}
    1,  & \text{if} v=v' \\
    \nu\in (0,1), & \text{otherwise}.
  \end{cases}
\end{equation}
Here $\nu$ is a hyperparameter that will be learned using the training data set.
The edge elementary kernel is a square exponential kernel (i.e., Gaussian
kernel) function on edge lengths, which is $1$ if two edges are of the
same length, and it smoothly changes to $0$ as the difference in lengths grows:
\begin{equation}
  \label{eq:e_kernel}
  K_e(e,e') = \exp\left[-\dfrac{1}{2} \dfrac{(e-e')^2}{\lambda^2}\right].
\end{equation}
The adjacency rule that computes the weights for each edge also assumes a square
exponential form
\begin{equation}
  \label{eq:adj}
  \mathcal{A}(\bm r_i, \bm r_j) =
  \exp\left[-\dfrac{1}{2}\dfrac{\Vert\bm r_i-\bm r_j\Vert^2}{(\zeta\sigma_{ij})}\right]
\end{equation}
where $\sigma_{ij}$, are element-wise length scale parameters derived from
typical bonding lengths. A uniform starting probability $p_s(\cdot)\equiv s$ and
a uniform stopping probability $p_q(\cdot)\equiv q$ are used across all
vertices.

Given a training set $D$ of $m$ molecules and their associated solvation free energy $\{(M_1,\cdots,M_m)\}$, $\{(E_1,\cdots, E_m)\}$, as well as a marginalized graph kernel $K(\cdot,\cdot)$, the GPR prediction for the energy $\{E^*_1,\cdots, E^*_n\}$ of a test set of n unknown molecules $\{M^*_1,\cdots,M^*_n\}$ can be derived analytically as
\begin{equation}
\label{eq:graph_pred}
    \bm E^*\defeq [E^*_1,\cdots,E^*_n]^\trans = \bm K_{D^*} \bm K_{DD}^{-1} \bm y_D,
\end{equation}
and the uncertainty in the prediction is given as:
\begin{equation}
\label{eq:graph_var}
    \bm\Sigma^*\defeq \bm K_{**}-\bm K_{D^*}^\trans \bm K_{DD}^{-1} \bm K_{D^*}.
\end{equation}
Here, $K_{DD}$ is an $n\times n$ matrix with $K_DD(i,j)=K(M_i, M_j)$, $\bm K_{D^*}$ is an $n\times m$ matrix with $\bm K_{D^*}(i,j)=K(M_i, M_j^*)$ and $\bm K_{**}$ is an  $m\times m$ matrix with $\bm K_{**}(i,j)=K(M_i^*,M_j^*)$.

\subsection*{Details of GPR}
In the GPR method, the mean and covariance functions $\mu(\bm x)$ and 
$k(\bm x,\bm x')$ are obtained by identifying their hyperparameters via 
maximizing the log marginal likelihood~\cite{williams2006gaussian}:
%-------------------------------------------------------------------------------
\begin{equation}
  \label{eq:lml}
  \ln L=-\dfrac{1}{2}(\bm y-\bm\mu)^\trans \tensor C^{-1} (\bm
  y-\bm\mu)-\dfrac{1}{2}\ln |\tensor C|-\dfrac{N}{2}\ln 2\pi.
\end{equation}
%-------------------------------------------------------------------------------

Moreover, to account for the observation noise, one can assume that the noise is
independent and identically distributed (i.i.d.) Gaussian random variables with
zero mean and variance $\delta^2$, and replace $\tensor C$ with 
$\tensor C+\delta^2\tensor I$. In this study, we assume that observations 
$\bm y$ are noiseless. If $\tensor C$ is not invertible or its condition number
is very large, one can add a small regularization term $\alpha\tensor I$
($\alpha$ is a small positive real number) to $\tensor C$, % such that it becomes full rank. 
which is equivalent to assuming there
is an observation noise. In addition, $\hat s$ can be used in global
optimization, or in the greedy algorithm to identify locations of additional
observations. 

Given a stationary covariance function, the covariance matrix $\tensor C$ can be written as
$\tensor C=\sigma^2\tensor \Psi$, where
$\Psi_{ij}=\exp(-\frac{1}{2}\Vert\bm x^{(i)}-\bm x^{(j)}\Vert_w^2)$. The estimators of $\mu$ and $\sigma^2$, denoted as $\hat\mu$ and
$\hat\sigma^2$, are
%-------------------------------------------------------------------------------
\begin{equation}
\label{eq:krig_est}
   \hat\mu=\dfrac{\bm 1^\trans \tensor\Psi^{-1}\bm y}
                  {\bm 1^\trans\tensor\Psi^{-1}\bm 1}, \qquad
   \hat\sigma^2=\dfrac{(\bm y-\bm 1\hat\mu)^\trans\tensor\Psi^{-1}(\bm y-\bm 1\hat\mu)}{N},
\end{equation}
%-------------------------------------------------------------------------------
where $\bm 1$ is a constant vector consisting of $1$s~\cite{forrester2008engineering}. 
It is also common to set $\mu=0$~\cite{williams2006gaussian}. The hyperparameters $\sigma$ and $l_i$ are 
identified by maximizing the log marginal likelihood in Eq.~\eqref{eq:lml}. The terms
$\hat y(\bm x^*)$ and $\hat s^2(\bm x^*)$ in Eq.~\eqref{eq:krig} take the 
following form:
%-------------------------------------------------------------------------------
\begin{align}
\label{eq:krig_mean}
\hat y(\bm x^*) & = \hat\mu + \bm\psi^\trans\tensor\Psi^{-1}(\bm y-\bm 1\hat\mu), \\
\label{eq:krig_var}
\hat s^2(\bm x^*) & = \hat\sigma^2\left(1-\bm\psi^\trans\tensor\Psi^{-1}\bm\psi\right),
\end{align}
%-------------------------------------------------------------------------------
where $\bm\psi=\bm\psi(\bm x^*)$ is a (column) vector consisting of correlations between the observed data and the
prediction, i.e., $\psi_i=\frac{1}{\sigma^2}k(\bm x^{(i)},\bm x^*)$.
%-------------------------------------------------------------------------------

\subsection*{Details of GPR using graph kernel}

In Eq.~\eqref{eq:distance}, the straightforward enumeration is impossible, because $\ell$ spans from $1$ to $\infty$. Nevertheless, Eq.~\eqref{eq:distance} can be reformulated under the spirit of dynamic programming as follows:
\begin{equation}
    K(G, G')=\sum_{h_1\in V, h'_1\in V'}p_s(h_1)p_s'(h'_1)K_v(h_1, h_1')R_{\infty}(h_1,h_1'),
\end{equation}
where $R_{\infty}$ is the solution to the following (linear) equilibrium equation
\begin{equation}
 R_{\infty}(h_1,h_1')=p_q(h_1)p'_q(h_1')+\sum_{i\in V, j\in V'}t(i,j,h_1,h_1')R_{\infty}(i,j), \label{eq:rinf}
\end{equation}
where 
\begin{equation}
    t(i,j,h_1,h_1')\defeq p_t(i|h_1) p_t'(j|h_1')K_v(v_i,v_j)K_e(e_{ih_1},e_{jh_1'}).
\end{equation}

Equation \ref{eq:rinf} exhibits a Kronecker product structure, which can be readily recognized in matrix form~\cite{tang2019prediction}:
\begin{equation}
  \mathbf{r}_\infty = \mathbf{q} \otimes \mathbf{q}' + \left[ \left( \mathbf{P} \otimes \mathbf{P}' \vphantom{\overset{\kappa_\mathrm{v}}{\smallotimes}} \right) \odot \left( \mathbf{E} \overset{\kappa_\mathrm{e}}{\smallotimes} \mathbf{E}' \right) \right] \cdot \mathbf{diag}\left( \mathbf{v} \overset{\kappa_\mathrm{v}}{\smallotimes} \mathbf{v}' \right) \cdot \mathbf{r}_\infty, \label{eq:r-infinity-linear-system}
\end{equation}
where
\begin{itemize}[label=,leftmargin=8em]
  \item[$\mathbf{v}\phantom{'}$] is the vertex label vector of $G$ with $\mathbf{v}_i = v_i$;
  \item[$\mathbf{p}\phantom{'}$] is the starting probability vector of $G$ with $\mathbf{p}_i = p_s(v_i)$;
  \item[$\mathbf{q}\phantom{'}$] is the stopping probability vector of $G$ with $\mathbf{q}_i = p_q(v_i)$;
  \item[$\mathbf{P}\phantom{'}$] is the transition probability matrix of $G$ defined as $\mathbf{D}^{-1} \mathbf{A}$;
  \item[$\mathbf{E}\phantom{'}$] is the edge label matrix of $G$ with $\mathbf{E}_{ij} = e_{ij}$;
  \item[$\mathbf{v}'$, $\mathbf{p}'$, $\mathbf{q}'$, $\mathbf{P}'$, $\mathbf{E}'$] are the corresponding vectors and matrices for $G'$;
  \item[$\overset{\kappa_\mathrm{v}}{\smallotimes}\phantom{'}$] is the generalized Kronecker product between $\mathbf{v}$ and $\mathbf{v}'$ with respect to microkernel $\kappa_\mathrm{v}$;
  \item[$\overset{\kappa_\mathrm{e}}{\smallotimes}\phantom{'}$] is the generalized Kronecker product between $\mathbf{E}$ and $\mathbf{E}'$ with respect to microkernel $\kappa_\mathrm{e}$.
\end{itemize}

\subsection*{Machine learning model}
\begin{figure}
    \centering
    \includegraphics[width=1\textwidth]{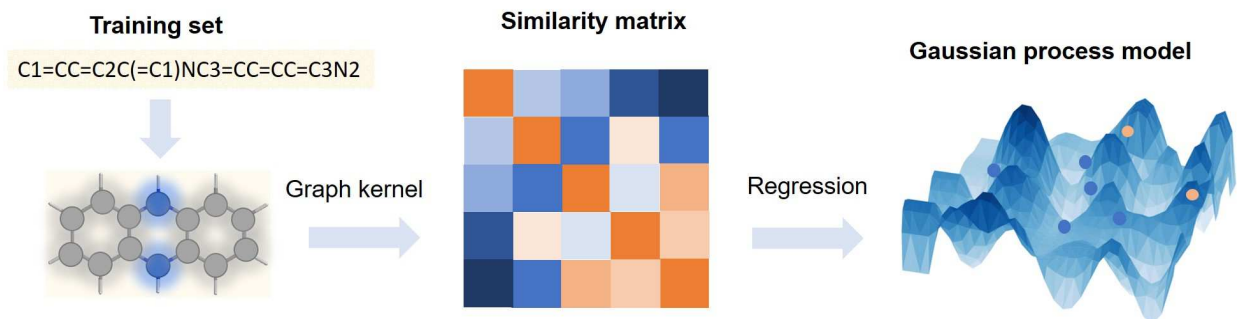}
    \caption{Scheme of the machine learning model pipeline}
    \label{fig:P1}
\end{figure}

Figure \ref{fig:P1} presents a scheme of the predictive machine learning model framework by Gaussian process regression with graph kernel. First, the SMILES string of molecules in the data set are converted to graph, where the atoms are the nodes and the bonds are the edges. The graph kernel is then applied to average over the similarities of all paths generated from simultaneous random walks on each pair of graphs. A predictive model with Gaussian process regression can be built by the pairwise similarity matrix among the training molecules and the cross-similarity matrix between the new molecule and the training molecules.
%Further details are available in the Methods section.

\subsection*{Metrics} 
In order to compare with the results, in this paper, mean absolute error (MAE) and root mean square error (RMSE) are applied to evaluate the performance of the ML model on the regression tasks. 
\begin{equation}
\label{eq:mae}
   \text{MAE}=\dfrac{1}{n}{\sum_{n=1}^{n}{\lvert \hat{y_i}-y_i \rvert}}.
\end{equation}

\begin{equation}
\label{eq:rmse}
   \text{RMSE}=\sqrt{\frac{1}{n}\Sigma_{i=1}^{n}{\Big({\hat{y_i} -y_i}\Big)^2}}.
\end{equation}
where $n$ is the number of molecules, ${y_i}$ is the solvation free energy value in database, $\hat{y_i}$ is the prediction solvation free energy by the ML model.

\subsection*{Cross-Validation and Hyperparameter Optimization}
%Since the data sets are not very large, 
We use the standard cross-validation approach to help identify the hyperparameters in the ML model, i.e., to perform model selection. For consistency, we maintain the same approach for all of our data sets. Specifically, for each data set, we split the data into training-validation and testing parts as described in Section~\ref{subsec:data}. 
%One of them is used for test set. The other nine parts are used as training-validation set to determine the best hyperparameters as shown above. 
%We select hyperparameters based on validation performance. 
%We then evaluate the model by retraining the model using the optimal hyperparameters and total train-validation set as new training set, and its performance is checked on the test set.
We employ 10-fold cross-validation (CV) for secure representation of the test data because the data set has a limited number of measurements. The molecules in the training-validation set of each data set is further split into 10 subsets following the sequence (InChIKey) of molecules. We choose one of the subsets as a validation set iteratively. The training set is the sum of the remaining 9 subsets. Consequentially, a 10-fold CV task performs 10 independent training and validation runs, and relative sizes of the training and validation sets are 9 to 1. 
We use Scikit-Learn library to implement the CV task and perform an extensive grid search for tuning hyperparameters. The hyperparameter set is determined by the result which has the minimum averaged MAE in the 10-fold CV. All the training is performed using our GPU-accelerated graph-kernel GPR tool\cite{tang2019prediction}.

\section{Results and Discussion}
%\section*{Results}
%In thermodynamics, the solvation free energy can be considered as a combination of the intramolecular electrostatic interaction (solvation energy), the nonpolar interaction contribution of solvent, and the contribution of conformational entropy. In this work, to characterize these effects on data in a machine learning model, three data sets obtained from DFT calculation, molecular dynamics (MD) simulation, and experiment with implicit or explicit water solvent were tested. 
%In addition, fordifferent data sets, we also experimentally determine the hyperparameters of the model.

\subsection*{Database}
\label{subsec:data}
In order to test the performance of the model on the prediction of solvation free energy, three data sets are built. Data set A1 is the solvation energy data obtained from DFT calculation with implicit water model. The molecules are selected from our own database. This solvation energy data set has 3626 molecules. All the molecules in the data set are neutral organic molecules. These molecules in the data set include ten types of elements, i.e., C, H, O, N, P, S, F, Cl, Br and I. All the solvation energy data in the data set are obtained from DFT calculation by PBE0 functional\cite{Perdew1996jcp} at 6-31G** level\cite{Ditchfield1971jcp} at 298.15K with NWChem code\cite{apra2020jcp}. An effect of implicit water solvent with a dielectric constant of 78.4 is included via the COnductor like Screening MOdel for Real Solvents (COSMO) model.\cite{Klamt1995jpc} These molecules are split into two sets as the training-validation set and test set following the sequence of their International Chemical Identifier key (InChIkey). Finally, 3200 molecules are selected in the training-validation set and 426 molecules are in the test set. 
Data set B1 is the solvation free energy data calculated by MD simulation in implicit water model. These data are obtained from a recently published paper. \cite{Rauer2020jcp} The original molecules are chosen from the QM9 database. QM9 consists of 134k molecules with up to nine heavy atoms, including chemical elements C, H, O, N, and F. In this data set, molecules containing fluorine are removed by the authors. They randomly selected 4000 compounds from the QM9 database and calculated their solvation free energy by MD simulation with implicit water model. However, after carefully examining the InChikey of these molecules, we find 24 duplicates in the database. Therefore, we only select data from 3976 molecules from this database. Finally, 3600 molecules are used in the training-validation set and 376 molecules are in the test set.
Data set C1 is obtained from the Freesolv database, which includes the solvation free energy both in experiment and MD simulation with explicit water model as solvent.\cite{DRM2017jced} The experimental solvation free energy data are selected as our target in this work. To keep consistent with the other two databases, we do not use the solvation free energy data of chiral molecules in the Freesolv database. After excluding the chiral molecules, we select 588 molecules. The molecules in this database also include ten elements, i.e., C, H, O, N, P, S, F, Cl, Br and I. The 588 molecules are divided into two sets. The training-validation set includes 550 molecules and the test set has 38 molecules. Figure 1 shows the probability distribution function (PDF) of the training-validation set and test set for the three data sets. We can see that the train-validation set and test set in each data set have similar PDFs of solvation free energy. As the size of data set C1 is smaller, the fluctuation in the PDF is stronger than the other two databases. Overall, Figure 1 indicates that it is reasonable using the identifier InChIkey for random splitting data, especially when the data set is not very small, e.g., larger than one hundred molecules. In the ML model building, We use a Simplified Molecular Input Line Entry System  (SMILES) string as initial input identifier in this work. The SMILES strings of molecules are converted to a graph with our graphic kernel when building ML models. 
\begin{figure}[t]
%    \centering
    \subfigure[A1]{
    \includegraphics[width=0.48\textwidth]{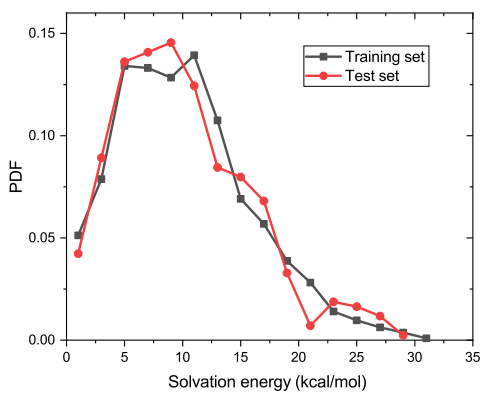}}
    \subfigure[B1]{
    \includegraphics[width=0.48\textwidth]{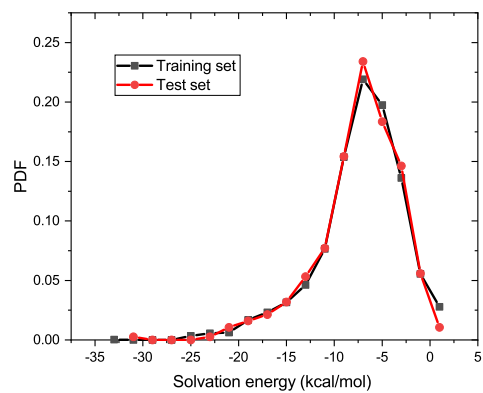}}
    \subfigure[C1]{
    \includegraphics[width=0.48\textwidth]{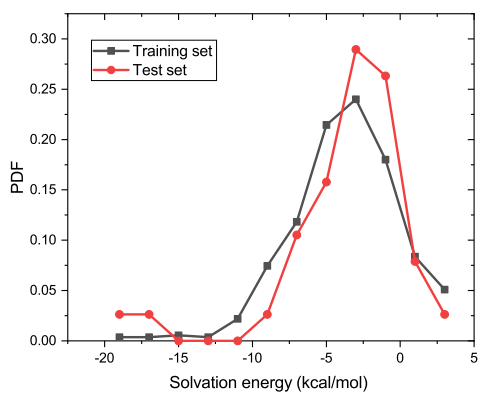}}
    \caption{Probability distribution function of solvation free energy in training data set and test data set of the three data sets.(a) A1. (b) B1. (c) C1.}
    \label{fig:pdf1}
\end{figure}

\subsection*{Solvation free energy prediction} 
\begin{figure}[!htb]
%    \centering
    \subfigure[Parity plot of data set A1]{
    \includegraphics[width=0.48\textwidth]{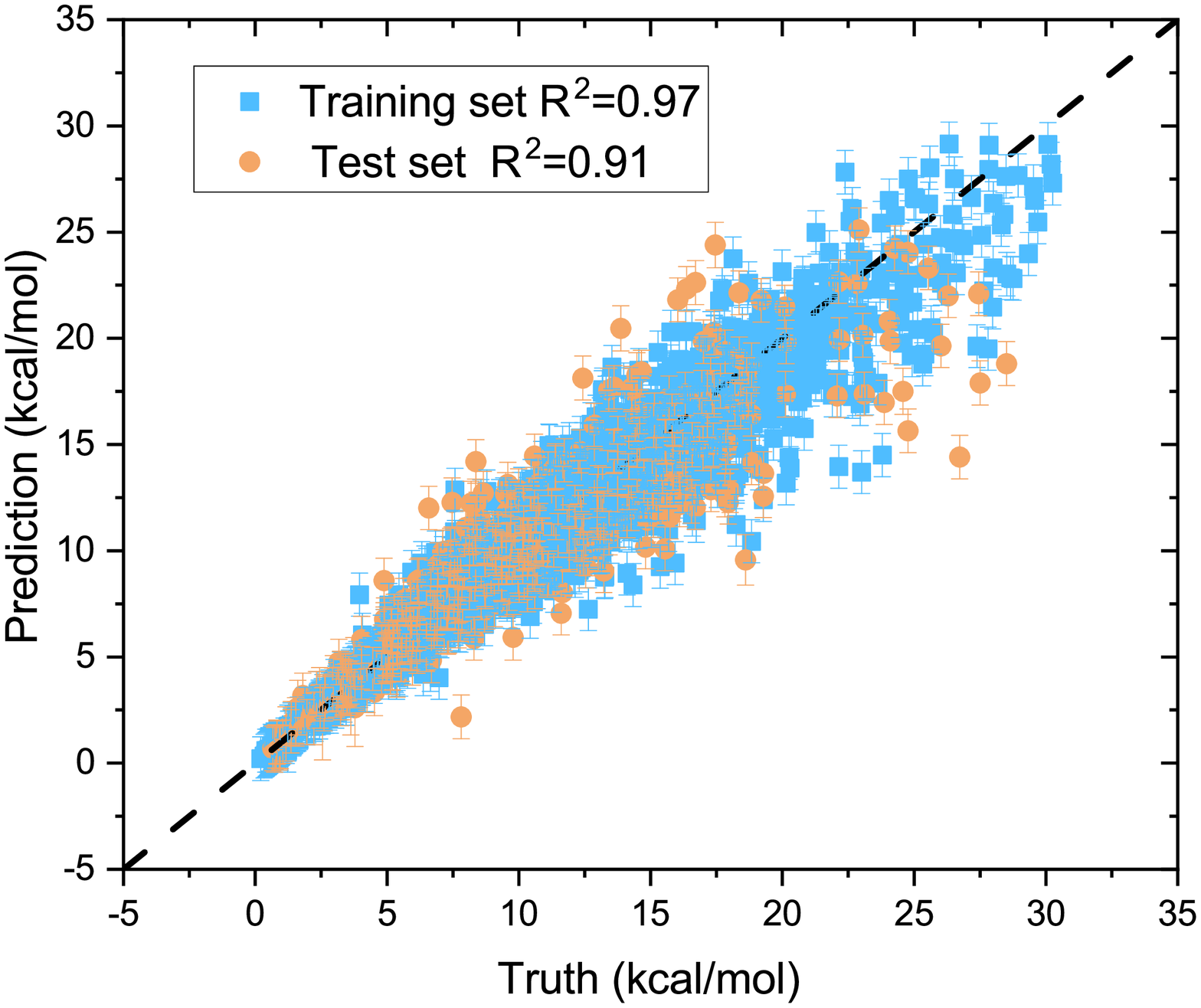}}
    \subfigure[Parity plot of data set B1]{
    \includegraphics[width=0.48\textwidth]{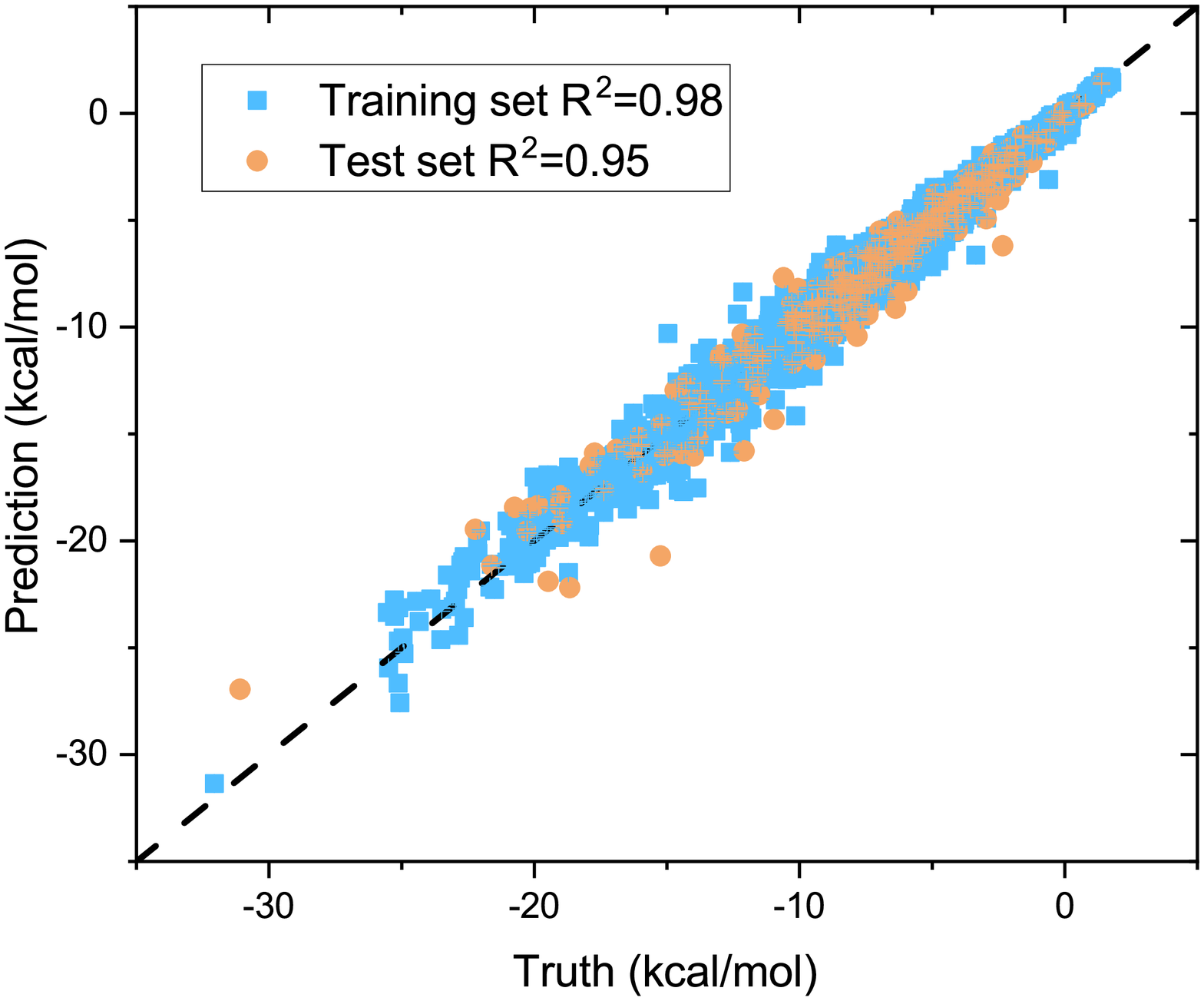}}
    \subfigure[Parity plot of data set C1]{
    \includegraphics[width=0.48\textwidth]{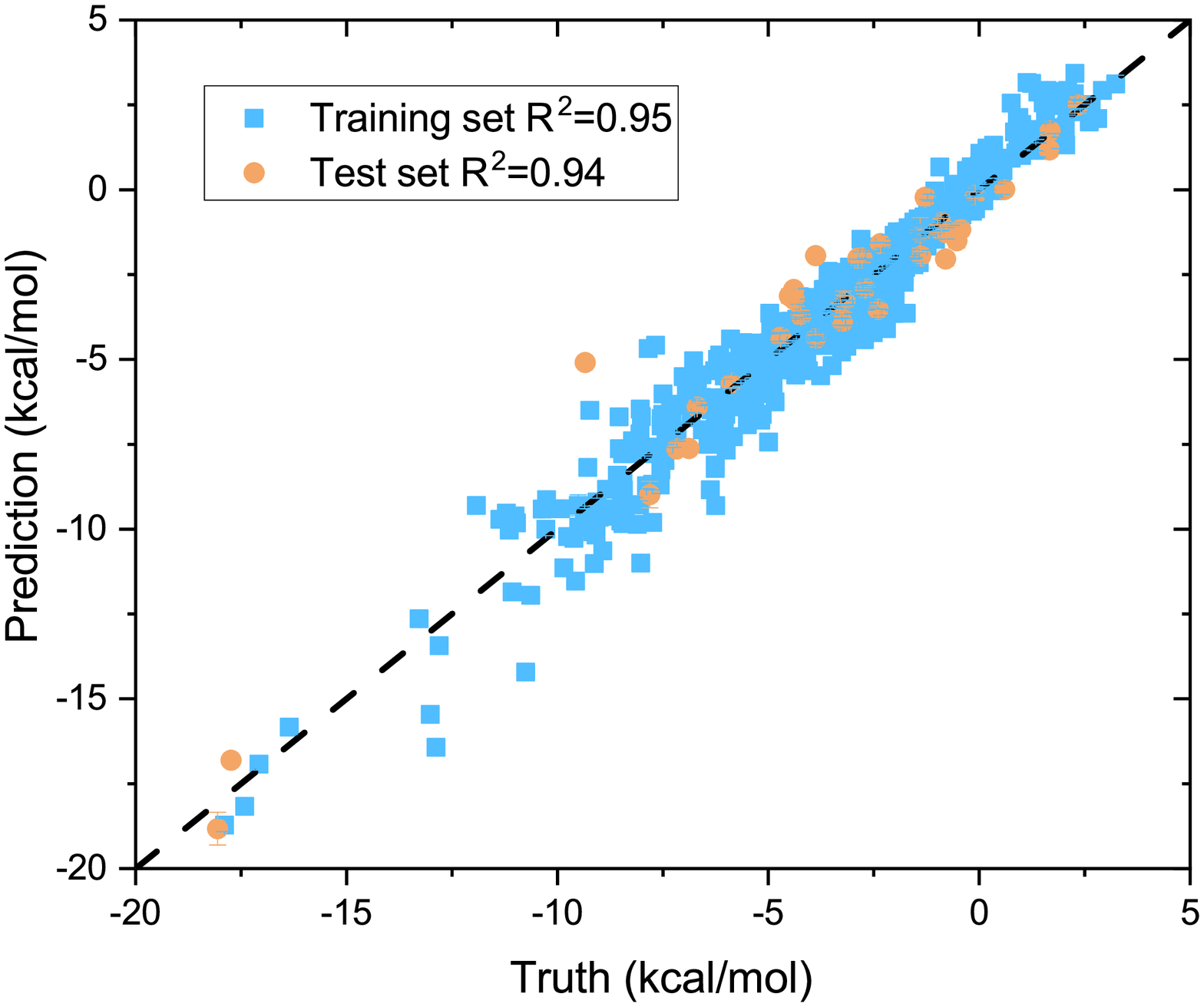}}
    \subfigure[MAE and RMSE in data sets A1]{
    \includegraphics[width=0.48\textwidth]{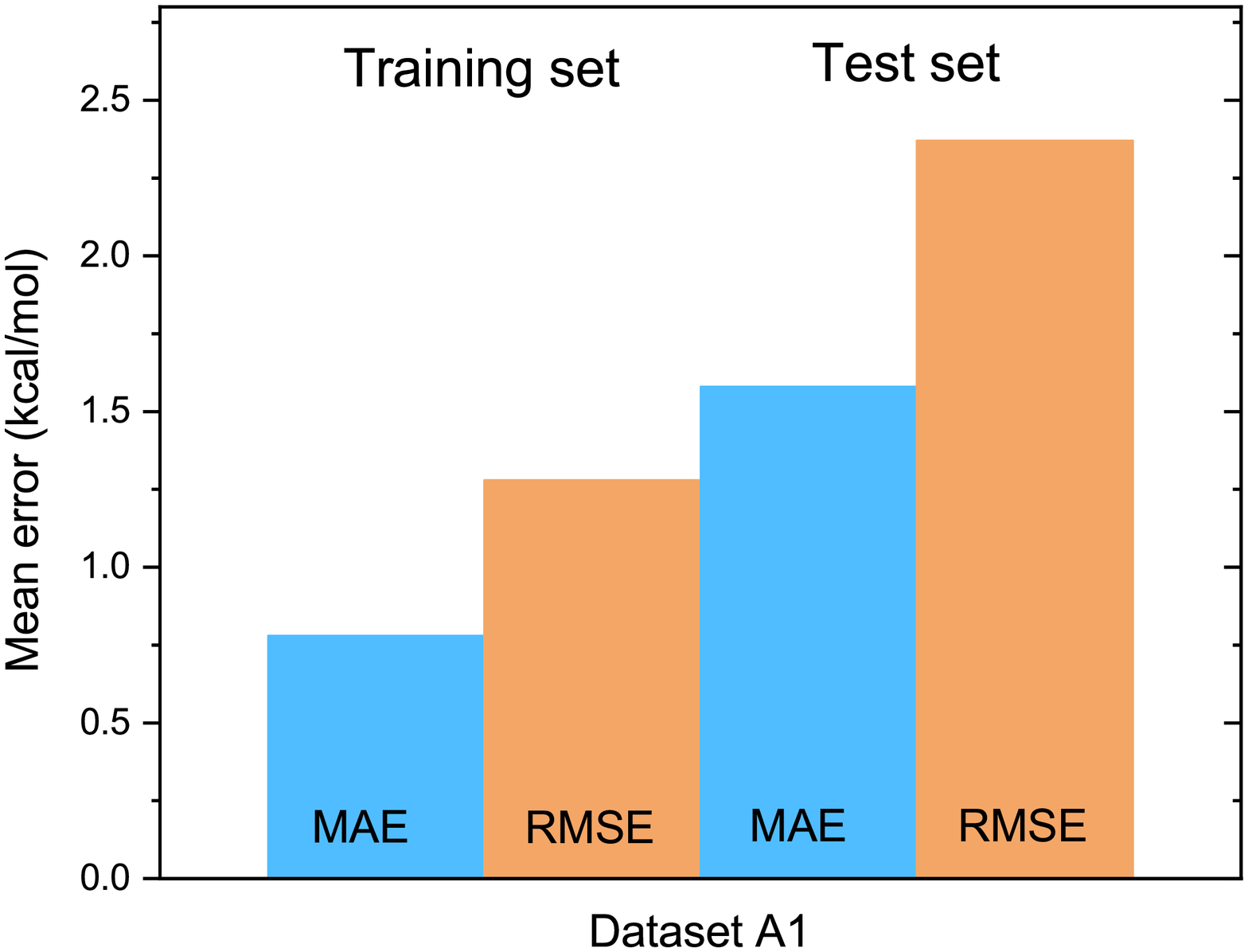}}
    \subfigure[MAE and RMSE in data sets B1]{
    \includegraphics[width=0.48\textwidth]{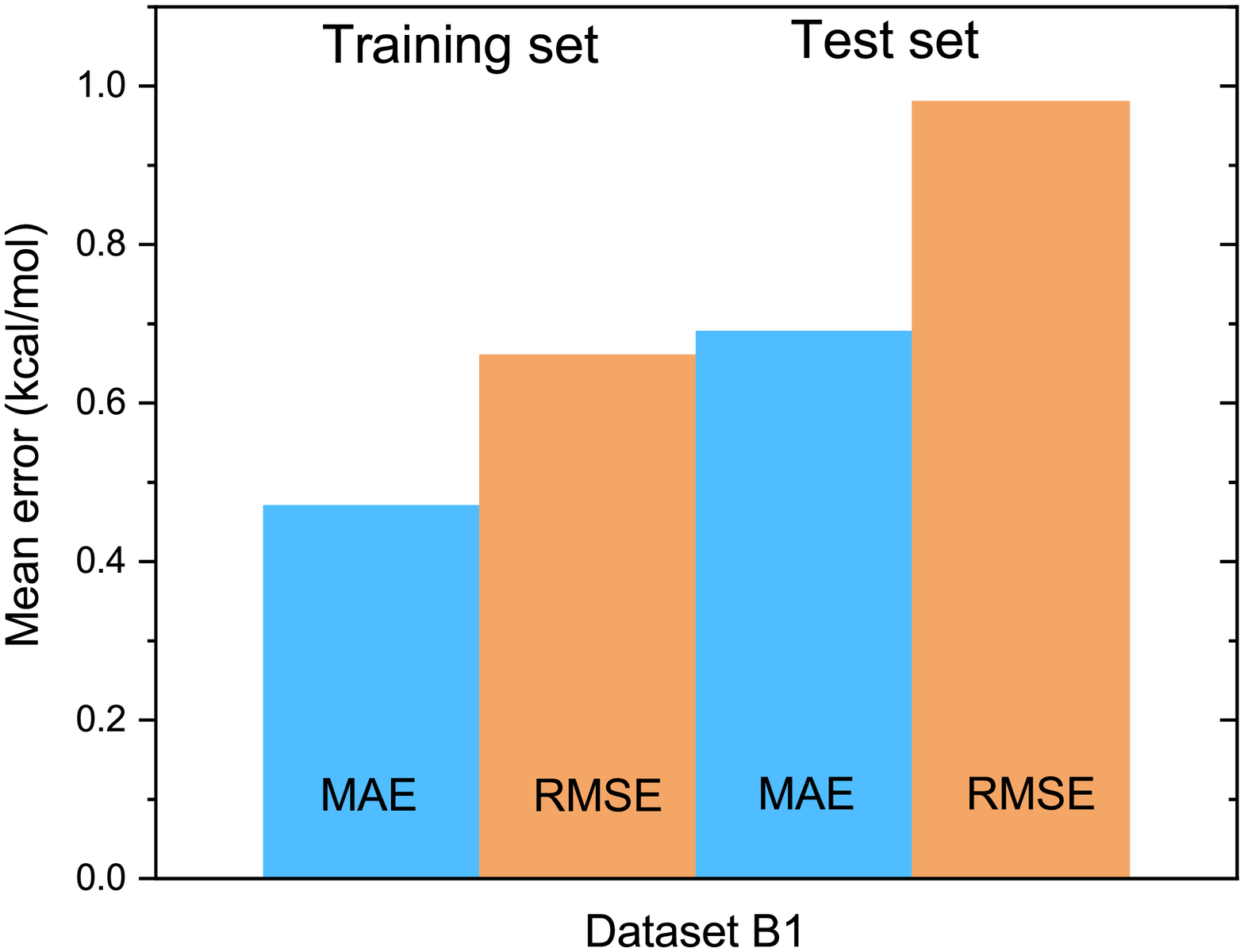}}
    \subfigure[MAE and RMSE in data sets C1]{
    \includegraphics[width=0.48\textwidth]{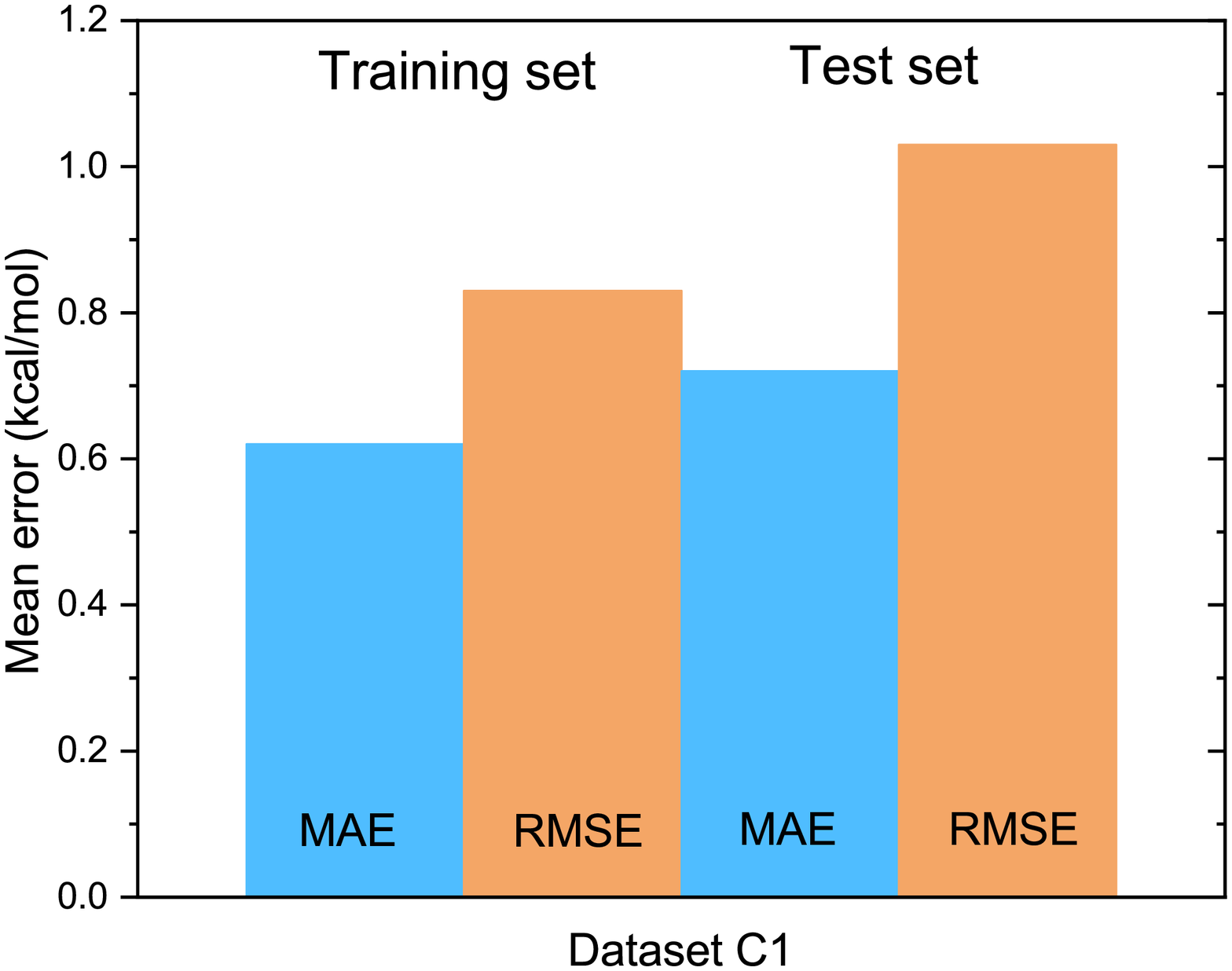}}
    \caption{Parity plots, MAE and RMSE of training data and test data in data sets A1, B1 and C1.}
    \label{fig:ds1}
\end{figure}

Solvation energies prediction results of the three data sets are displayed in Figure~\ref{fig:ds1}. With the help of optimized hyperparameters, the results of the three data sets show good performance for our ML model in general. The Pearson correlation coefficients $R^{2}$ between the truth and the prediction for the training set in the three data sets are 0.97, 0.98 and 0.95, respectively. The $R^{2}$ of the test set in these three cases are 0.91, 0.95 and 0.94, respectively. We can see the Pearson correlation coefficients are in good agreement for training data and test data in each data set, implying our ML model is not overfitted. 
The results in Figure~\ref{fig:ds1} show that the predication accuracy for data sets B1 and C1 are better than for A1. The results are interesting, since in fact the measurement uncertainties of solvation free energy for the three data sets are increasing from A1 to C1. For DFT calculation, the measurement uncertainty for fixed functional and basis should be very small, as during the calculation the molecular conformation is fixed, and there is no thermal fluctuation. Therefore, the uncertainty should be \textless 0.01 kcal/mol. In MD simulation with implicit solvent model, due to the conformational change in MD simulation, the fluctuation of calculated solvation free energy is larger than the DFT calculation, which increases measurement uncertainty. In experiments, the uncertainty can be even larger than the MD simulation, which has been demonstrated in the Freesolv database. In the Freesolv database, the average error is about 0.06 kcal/mol for MD simulation data of solvation free energy, but for the experiment data it is 0.3 kcal/mol. However, by adding appropriate strength of white noise in the training process, we find that the uncertainty does not affect the accuracy of our ML models.
%Figure S1 shows that the relationship between the strength of white noise and the MAE in the training set and test set. We see that it is necessary to find an appropriate value of white noise to reduce the MAE and avoid overfitting. 
Note that in general, it is necessary to include an appropriate level of measurement error, i.e., noise, to avoid overfitting when training ML models. In the GPR model, as indicated in Section~\ref{subsec:gpr}, the noise is included in the covariance matrix.
If the noise level included in the ML model is too small, the model is prone to overfitting. If it is too large, the error in prediction would be also large. So noise is an important hyperparameter in the model parameterization. 

%\begin{figure}[t]
%    \centering
%    \subfigure[A1]{
%    \includegraphics[width=0.48\textwidth]{./figures/sol3200MAE.eps}}
%    \subfigure[B1]{
%    \includegraphics[width=0.48\textwidth]{./figures/qm9MAE.eps}}
%    \subfigure[C1]{
%    \includegraphics[width=0.48\textwidth]{./figures/freesolvMAE.eps}}
%    \includegraphics[width=0.5\textwidth]{./figures/sol3600dataPDF.pdf}
%    \caption{MAE and RMSE of training data and test data in datasets A1, B1 and C1. (a)A1. (b)B1. (c)C1.}
%    \label{fig:mae1}
%\end{figure}

Figure \ref{fig:ds1}, parts d-f present the MAE and RMSE in training set and test set for the three data sets. For MAE results in both training set and test set in each data set, the results are very close, indicating our ML model is not overfitted. The RMSE results also show the same trend as MAE in each data set, which verifies our conclusion. %Interestingly, the predicted solvation energy in data set A1 (DFT calculation result) has the largest MAE, which is 0.78 kcal/mol, although as discussed above, the theoretical uncertainty is the smallest in the three data sets. The RMSE is 1.28 kcal/mol, which is also the largest one among the three training sets. 
For the training set in data set A1, the MAE is 0.78 kcal/mol and the RMSE is 1.28 kcal/mol.
With regard to the test set in data set A1, the MAE and RMSE are close to the training set results but a little higher. The results are 1.58 kcal/mol and 2.37 kcal/mol, respectively. 
%The performance of our model on data set B1 is better than data set A1. 
For the data set B1, the MAE and RMSE are 0.47 kcal/mol and 0.66 kcal/mol for training set. The test set follows the same trend. The MAE and RMSE are 0.69 kcal/mol and 0.98 kcal/mol. For data set C1, the MAE and RMSE result are close to the result obtained in data set B1. The MAE and RMSE in the training set are only a little higher than in B1. They are 0.62 kcal/mol and 0.83 kcal/mol. The test set results are similar, 0.72 kcal/mol and 1.03 kcal/mol, respectively. 

It is a bit difficult to directly compare our results with other ML models because we either have different data sets or use a different split method for the data set. While we know that the error of energy in a DFT calculation with different functional/basis would be several kilo calories, from the above results We can see that our ML model has yield chemical accuracy (1 kcal/mol) for the QM9 database subset and Freesolv database. Therefore, the mean absolute error in our ML model is actually close or even better than the DFT calculation. For the QM9 database subset, the authors previously obtained MAE = 0.7 kcal/mol with 2500 molecules in the training set,\cite{Rauer2020jcp} while the MAE of our training set is 0.47 kcal/mol with 3600 training data. For Freesolv database, Wu et al. provided a benchmark study of 642 molecules with different QSPR/ML models.\cite{Wu2018cs} The range of RMSE obtained with different ML methods is from 1.15 to 2.05 kcal/mol. In Lim and Jung's paper they obtained RMSE = 1.19 kcal/mol.\cite{lim2019cs} Our RMSE result is 1.03 kcal/mol with the same but even smaller training set. These results suggest that our graphic GP model guarantees considerably good performance.

\begin{figure}[!htb]
    \centering
    \subfigure[The average BCI for each dataset]{
    \includegraphics[width=0.48\textwidth]{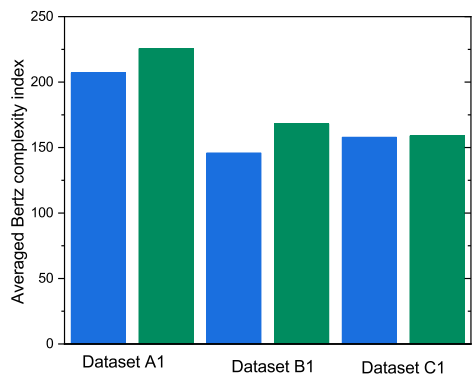}}
    \subfigure[The PDF of BCI for A1]{
    \includegraphics[width=0.48\textwidth]{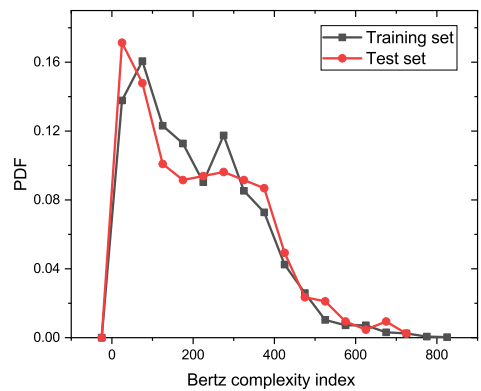}}
    \subfigure[The PDF of BCI for B1]{
    \includegraphics[width=0.48\textwidth]{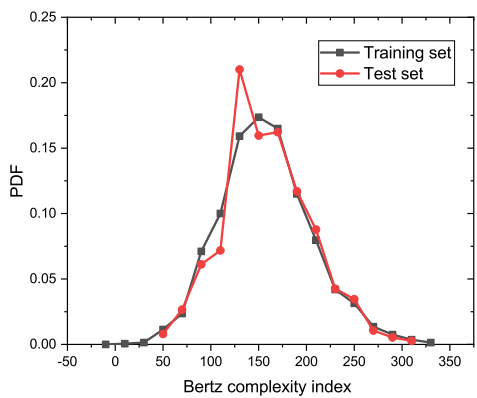}}
    \subfigure[The PDF of BCI for C1]{
    \includegraphics[width=0.48\textwidth]{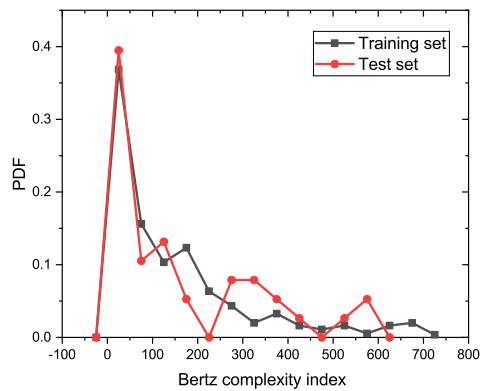}}
    \caption{The average Bertz complexity index and PDFs of Bertz complexity index for datasets A1, B1, and C1. blue bar, training set. green bar, test set.}
    \label{fig:aveb}
\end{figure}

The data set A1 has a large training set (3200 molecules), and theoretically the uncertainty of the data set A1 should be small. However, the performance of our model on data set A1 is not the best among the three data sets. For example, its $R^2$ is not the highest one of the data sets. One possible reason is that the complexity of this data set is higher. In data set A1 it involves ten types of elements. That means the converted molecular graph in data set A1 may have more types of nodes. In the view of graph theory, more types of nodes do not affect the topology, but they do increase the complexity of the molecular graph. Here, we use the Bertz complexity index to further characterize the complexity of the data set. The Bertz complexity index (BCI) \cite{bs1981jacs} is defined as following
\begin{equation}
\label{eq:bertz}
   \text{BCI}=2{n}\log_2{n}-{\sum_{l}{n_i\log_2{n_i}}},
\end{equation}
where ${n}$ is the number of pairs of adjacent edges in a graph G and ${n_i}$ is the number of pairs of adjacent edges in the ${i}$-th class by symmetry. 
The term ${n}\log_2{n}$ is used to prevent $\text{BCI} = 0$ when all pairs of adjacent edges in G are equivalent. We can see that the first part takes into account structural characteristics of G, such as size, branching, and cyclicity, and the second part deals with the symmetry of G in terms of equivalent pairs of adjacent edges. In other words, one represents the complexity of the bonding, the other represents the complexity of the distribution of heteroatoms. BCI has been used in analysis of synthetic strategies in organic chemistry \cite{bs1982jacs}, but it has not been connected to physical properties with the ML model. Figure \ref{fig:aveb}a shows the average BCI values of the three data sets. It is found that the average BCI of the training set and the average BCI of test set in each data sets are very similar. The average BCIs obtained from training set and test set in data set A1 are 207.0 and 220.5, respectively. For the other two data sets the BCI values are 157.9 and 158.9 in data set B1, and 145.9 and 168.1 in data set C1 for training set and test set, respectively. The data set A1 has the largest BCI. It implies that on average, the converted molecular graph in data set A1 is the most complicated. Therefore, more training data may be needed in order to reduce the MAE of the ML model on data set A1. The BCIs in data set B1 and C1 are close, although the type of elements in the two databases are not the same. It seems like the topological complexity in data set B1 and diversity of nodes in data set C1 have a complementary effect on BCI. 

To further investigate the effect of BCI on performance of the ML model, we calculated the PDFs of BCI for each data set. Figure \ref{fig:aveb} parts b to d present the PDFs of BCIs in each data set. It reveals more details of the data sets. In all three data sets, the PDFs of BCI for training set and test set are very close, which is similar to the PDFs of solvation free energy. That validates the split method of data set with InChikey is effective again. In addition, we identify that the shape of the PDFs for data set A1 and C1 are similar. They are both long-tailed distributions, like a Poisson distribution. That may be because more types of elements are included in these two data sets, as they both have ten elements. The peaks of these two PDFs are both between 0 to 50, which means the small molecules are main components in BCI, but the contribution of large molecules to the average BCI cannot be neglected. In data set A1, the contribution of large or complicated molecules in the tail part is higher than data set C1. That makes the final BCI larger in data set A1 than data set C1. For data set B1, its distribution is close to a Gaussian distribution. It does not include more molecules with high BCI as in the other two data sets. Thus, eventually, the data sets B1 and C1 have similar averaged BCIs. Also, as shown above, the predictions of our ML model on these two data sets are consistent with their complexity. Based on these results, we can infer that for a complicated data set like the molecular data set, the performance of a graphic ML model is not only related to the absolute amount of training data, but also the data complexity. As the dimension of molecular data may be quite high, that infers the data sparsity problem in high dimensional space for training data. 

%\subsection{Solvation free energy prediction in implicit solvent} 
%\subsection{Solvation free energy prediction in explicit solvent} 

%\begin{figure}[t]
%    \centering
%    \subfigure[A1]{
%    \includegraphics[width=0.48\textwidth]{./figures/sol3200bertzindexPDF.eps}}
%    \subfigure[B1]{
%    \includegraphics[width=0.48\textwidth]{./figures/qm9bertzindexPDF.eps}}
%    \subfigure[C1]{
%    \includegraphics[width=0.48\textwidth]{./figures/freesolvbertzindexPDF.eps}}
%    \includegraphics[width=0.5\textwidth]{./figures/sol3600dataPDF.pdf}
%    \caption{The PDFs of Bertz complexity index for datasets A1, B1, and C1. (a)A1. (b)B1. (c)C1.}
%    \label{fig:bpdf}
%\end{figure}

\begin{figure}[!ht]
    \centering
    \subfigure[A1]{
    \includegraphics[width=0.48\textwidth]{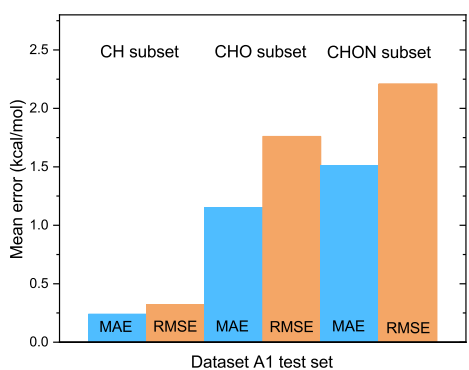}}
    \subfigure[B1]{
    \includegraphics[width=0.48\textwidth]{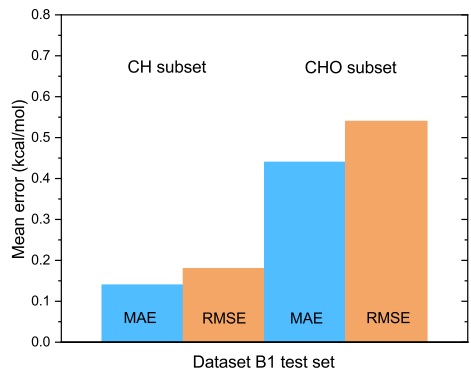}}
    \subfigure[C1]{
    \includegraphics[width=0.48\textwidth]{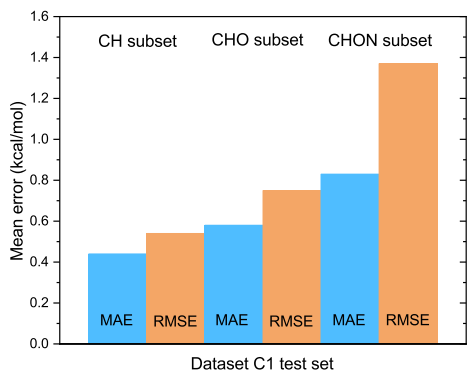}}
    \caption{MAE and RMSE of different subsets in test data of data sets A1, B1 and C1. (a) A1. (b) B1. (c) C1.}
    \label{fig:subset}
\end{figure}

For this reason, We do some tests with lower-dimensional subsets. We further evaluate the performance of our ML model with subsets in the test sets, which only include certain types of elements, e.g., C and H elements or C, H, and O elements. As shown in Figure \ref{fig:subset}, we see that all three data sets have the same trend. The MAE values increases with the element type complexity in these data sets. In these subsets, the simplest subset, which only includes the C and H elements, has the smallest MAE value. The MAE values are 0.24 kcal/mol, 0.14 kcal/mol, and 0.44 kcal/mol in data set A1, B1, and C1, respectively. These MAE values are much smaller than the MAE for the whole test set in these data sets. This is consistent with group contribution theory of solvation free energy, although the "groups" here are in high dimensional space. On the other hand, it indicates the ML model has relatively learned "more" information for compounds which only contain C and H elements from the training data. Additionally, we notice that the MAE value of the test group with C, H, O, and N elements in data set C1 is already higher than average in data set C1 test set (0.83 kcal/mol vs 0.72 kcal/mol), which implies the training data set is lacking molecules consisting of C, H, O, and N elements. The RMSE for the small test (1.37 kcal/mol) is also higher than the average value 1.24 kcal/mol. %{\color{blue}, which means such prediction is only because of a few molecules.}

\begin{figure}[!ht]
    \centering
    \subfigure[]{
    \includegraphics[width=0.48\textwidth]{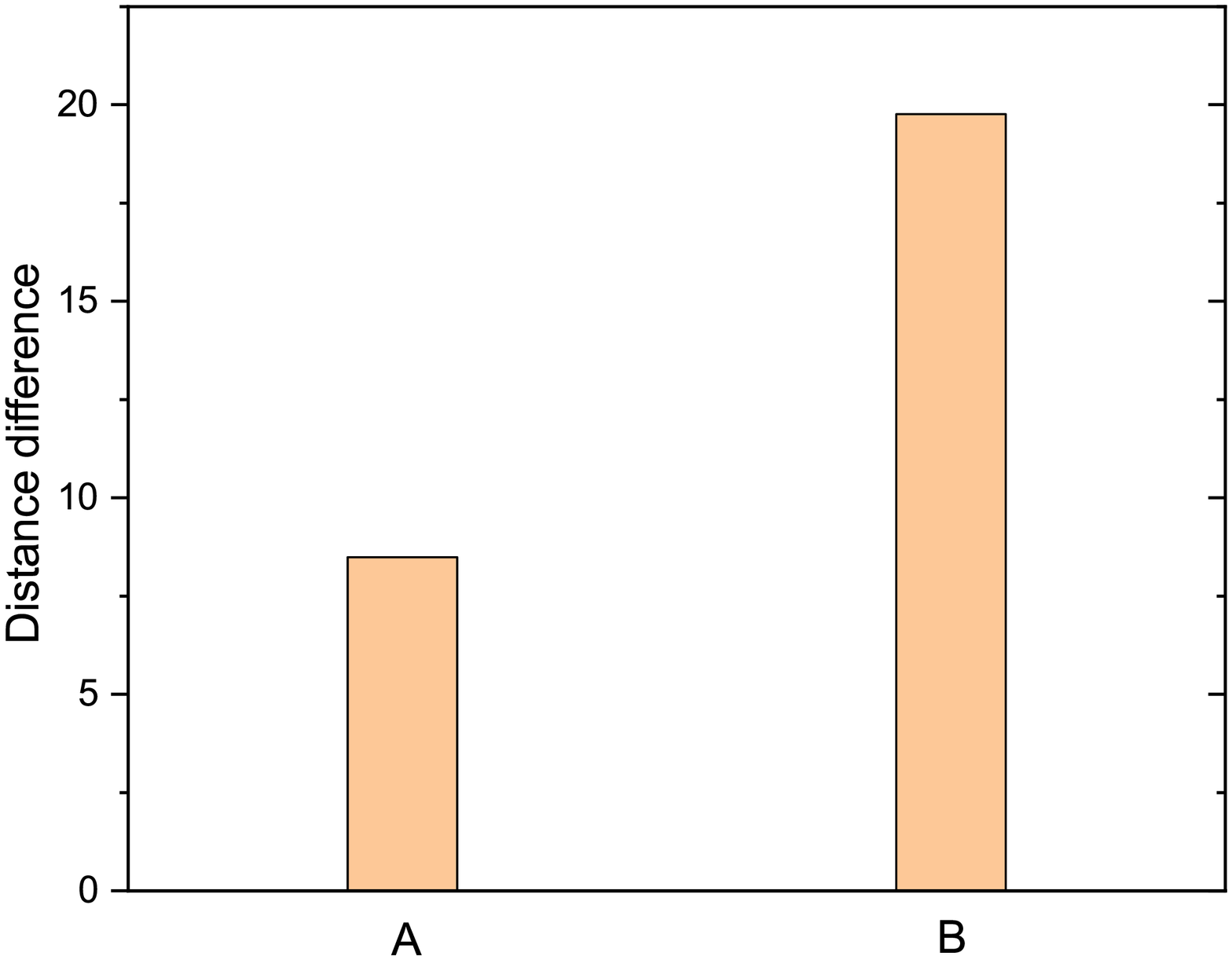}}
    \subfigure[]{
    \includegraphics[width=0.48\textwidth]{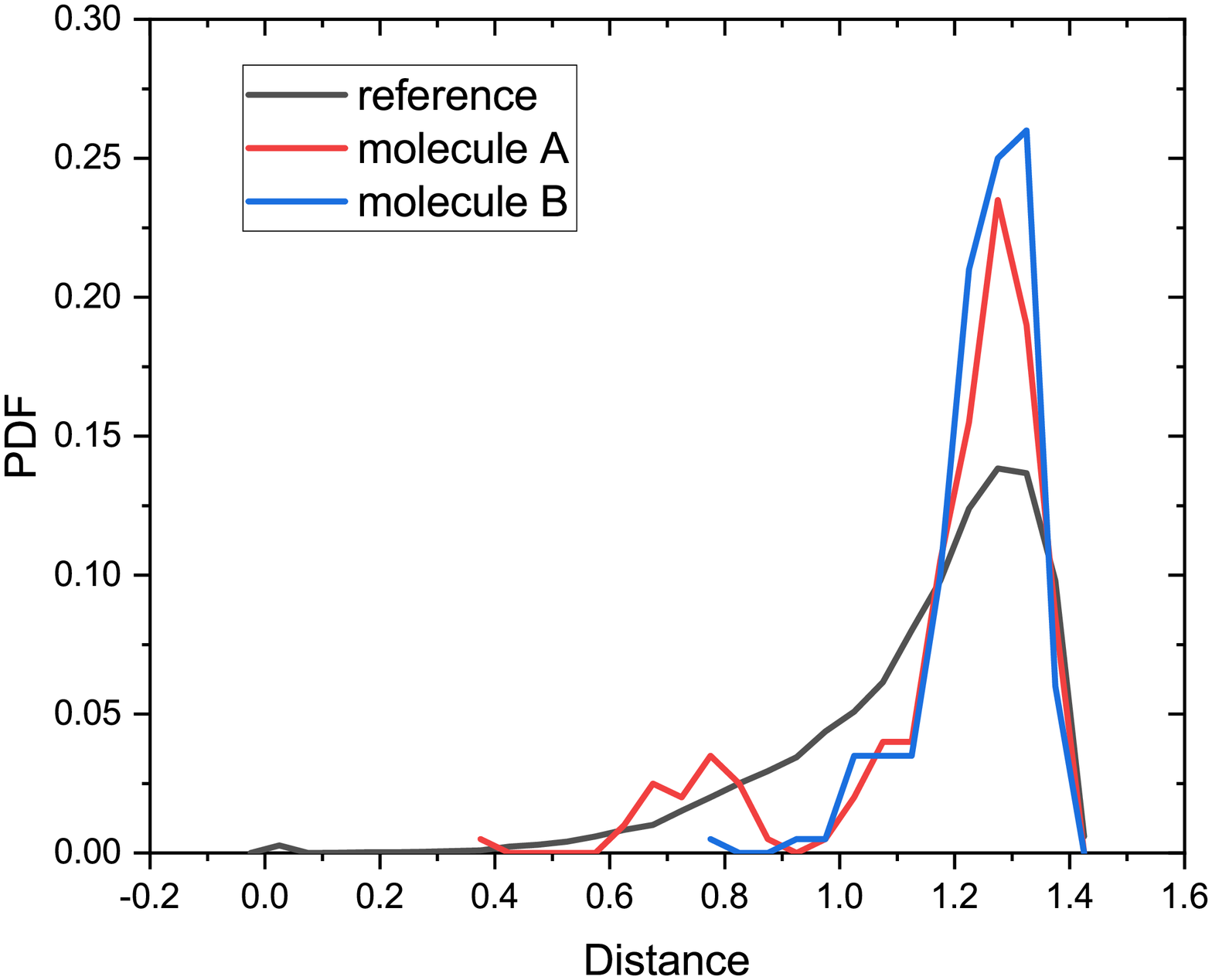}}
    \subfigure[]{
    \includegraphics[width=0.48\textwidth]{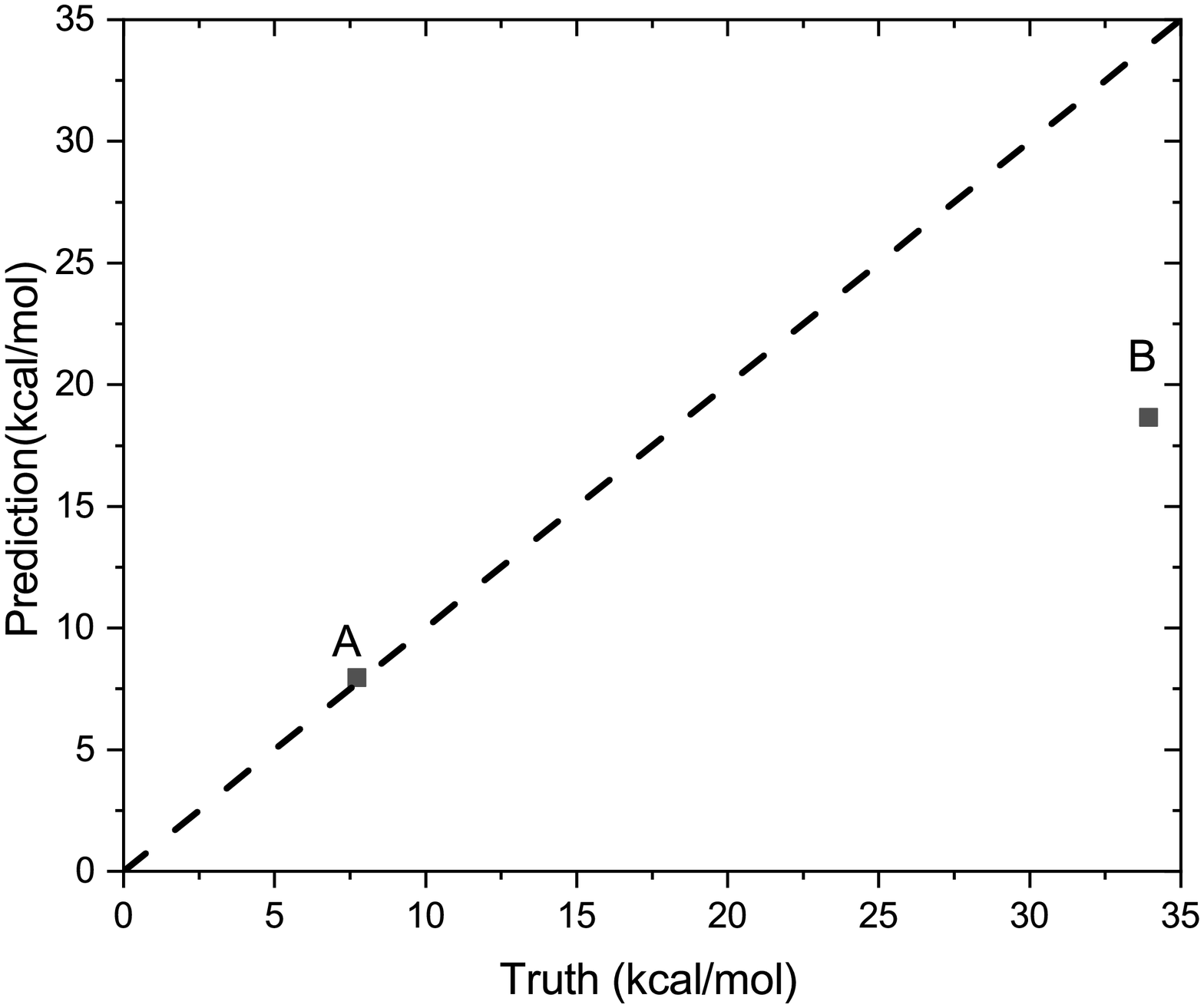}}
    \caption{Distances, PDF and prediction of two example molecules A and B. (a) Average distances between the training set and test molecules A and B. (b) PDF of pairwise distances between training molecules, distances between the training molecules and molecules A, and B, respectively. (c) The actual number and prediction for solvation energy of molecule A and B with the ML model.}
    \label{fig:smallexample}
\end{figure}

Additionally, we provide a method to qualitatively estimate performance of the ML model on predicting properties of new molecules via comparing the distances between molecular graphs in the test set and training set. Here we show an example of a subset with 200 molecules in data set A1 and select two molecules as the illustrative test set. We calculate average pairwise distances between molecules in the training set, and between the training set and each test molecule. The average distances in training set and each test molecule are displayed in Figure~\ref{fig:smallexample}(a). 
The PDFs of the distances are shown in Figure \ref{fig:smallexample}(b), which provides more details. We can find that the peak of PDF for molecule B is higher than molecule A, indicating the distance between the training set and B is farther than the distance between the training set and A in general. More importantly, the distances between molecule B and almost all training molecules are larger than 1.0, while there are some training molecules within the distance range of $[0.6,0.8]$ from molecule A.
Obviously, the distance for molecule A is much smaller than molecule B. In Figure~\ref{fig:smallexample}(c) we can also see the solvation energy prediction of molecule A is much better than molecule B. An important reason is that there are a sufficient number of training molecules that are close to molecule A, which results in a prediction with greater accuracy.

\subsection*{Dimension reduction}
To address the molecular data sparsity issue in high dimensional space and gain a deep understanding of the relationship between the training set and the ML model prediction, we analyze the training set with a model reduction approach. The covariance matrix that is used in the GP method plays a key role in the GPR, and it provides a possible way of exploring low-dimensional structures of the training data set that are critical to predict solvation free energy. In other words, it provides a possible way to identify critical functional groups (molecular fragments) that can be used as fundamental building blocks of real molecules, and the solvation free energy of a molecule can be predicted based on examining which groups are included in this molecule. To achieve this goal, we propose to associate molecules with points $Q_1, Q_2, \cdots, Q_m$ in Euclidean space $\mathbb{R}^d$, where $d$ is the dimension to be identified. We aim to use the distance matrix of the aforementioned points in $\mathbb{R}^d$ to approximate the covariance matrix, as such to identify an appropriate $d$. This $d$ is the number of the critical functional groups (or molecular fragments). Given a trained GP model and training data set, we have a covariance matrix $\tensor C$. For a fixed $d$, we generate points in $\mathbb{R}^d$ based on this $\tensor C$ as follows. We first define a matrix $\tensor T$ as 
\begin{equation}
T_{ij}=\dfrac{C_{1j}^2+C_{i1}^2-C_{ij}^2}{2}.    
\end{equation}      
Then we compute the eigenvalue decomposition of $\tensor T$:
\begin{equation}
    \tensor T=\tensor U \tensor S\tensor U^\trans.
\end{equation}
Finally, let $\tensor X=\tensor U\sqrt{\tensor S}$, and the first $d$ columns of $\tensor X$ are the desired $d$-dimensional points in $\mathbb{R}^d$. Of note, the distance matrix of $Q_i, i=1,2,\cdots,m$ generated in this way, denoted as $\tilde{\tensor C}$, is an approximation of the covariance matrix $\tensor C$ when $d<m$. Although it is possible that $\tilde{\tensor C}=\tensor C$, we can set a threshold for the difference $\Vert \tilde{\tensor C}-\tensor C\Vert_F$ to examine the accuracy of the approximation. Here $\Vert\cdot\Vert_F$ is the Frobenius norm of a matrix. 

\begin{figure}[!ht]
%    \centering
    \subfigure[A1]{
    \includegraphics[width=0.32\textwidth]{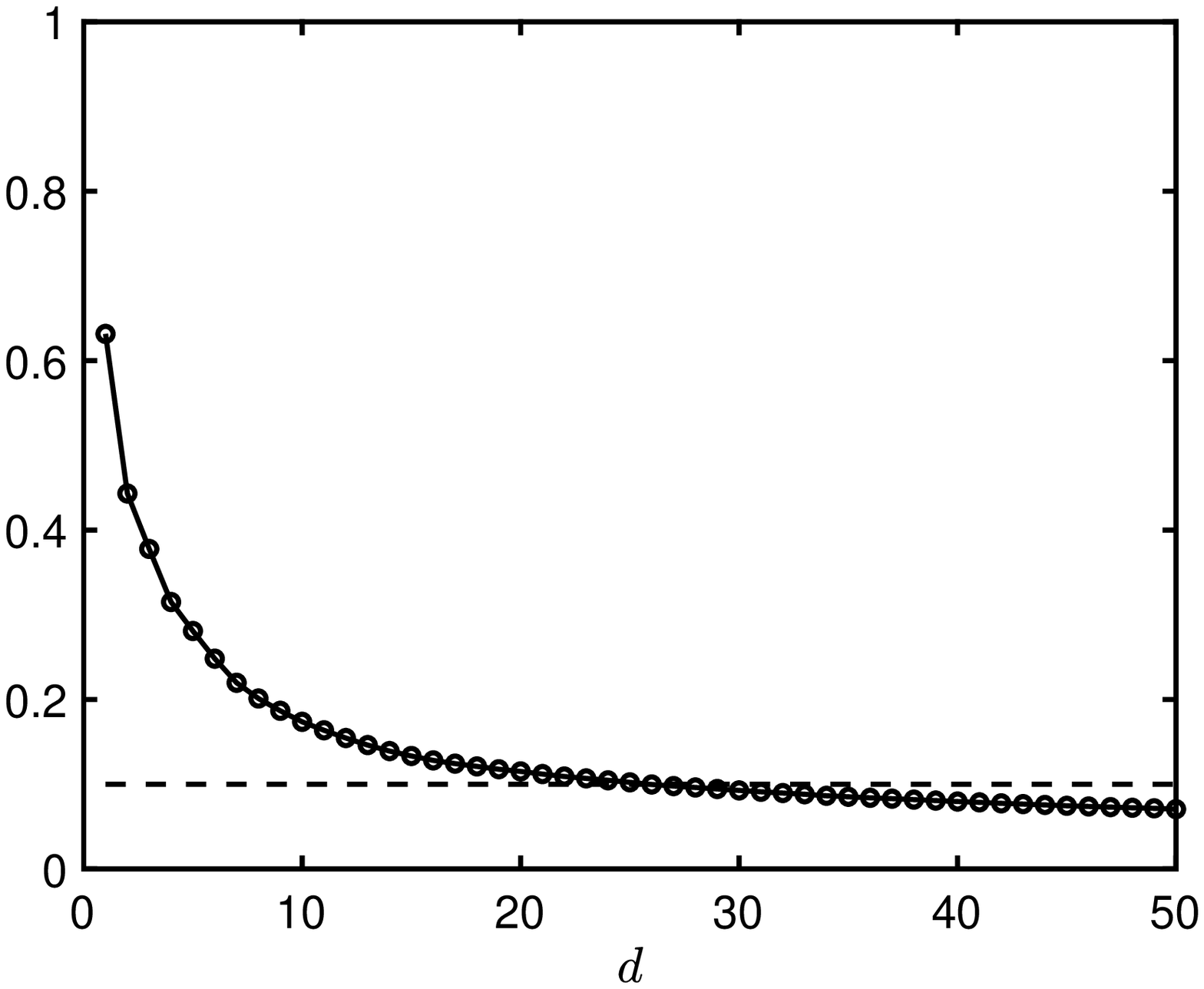}}
    \subfigure[B1]{
    \includegraphics[width=0.32\textwidth]{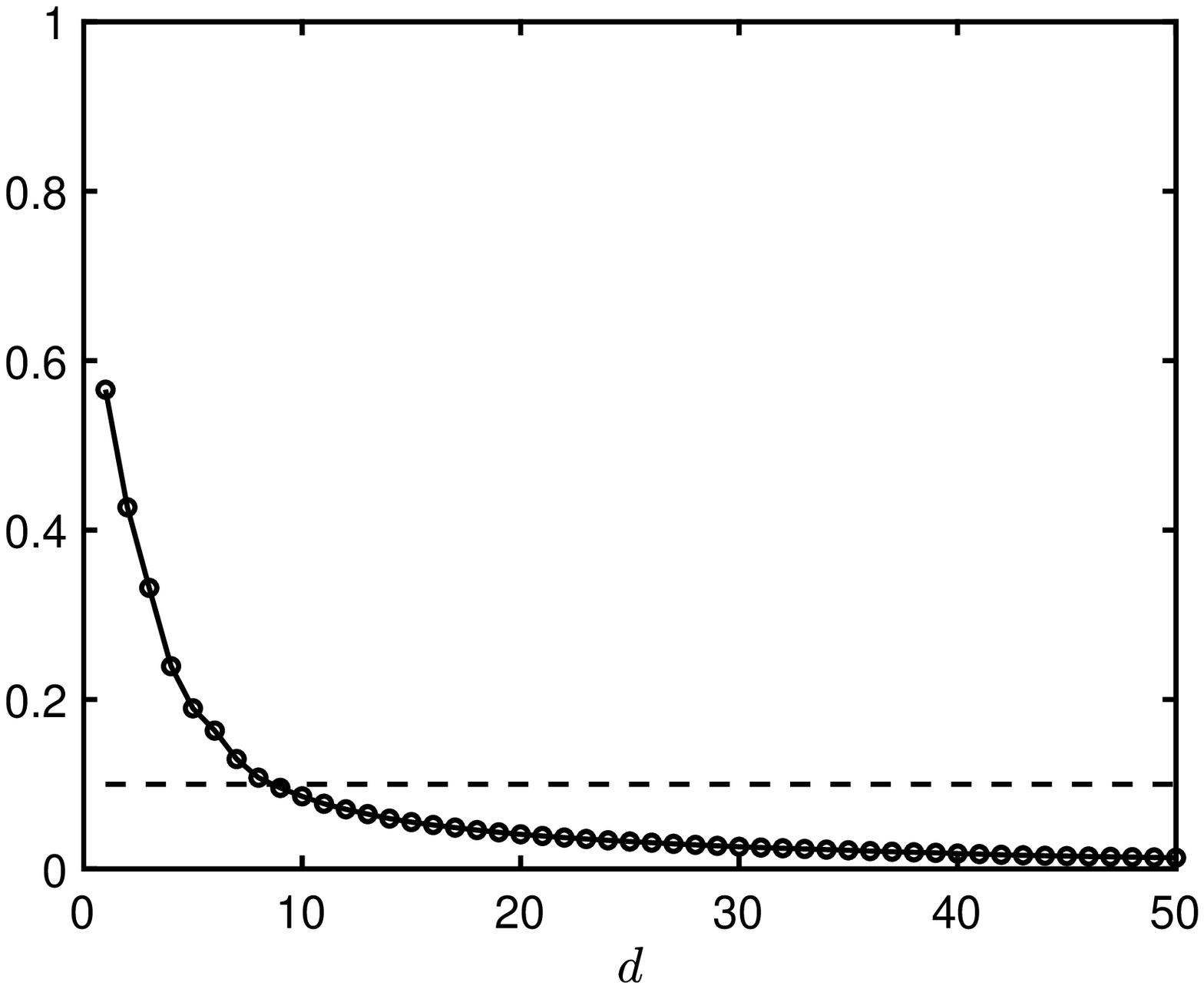}}
    \subfigure[C1]{
    \includegraphics[width=0.32\textwidth]{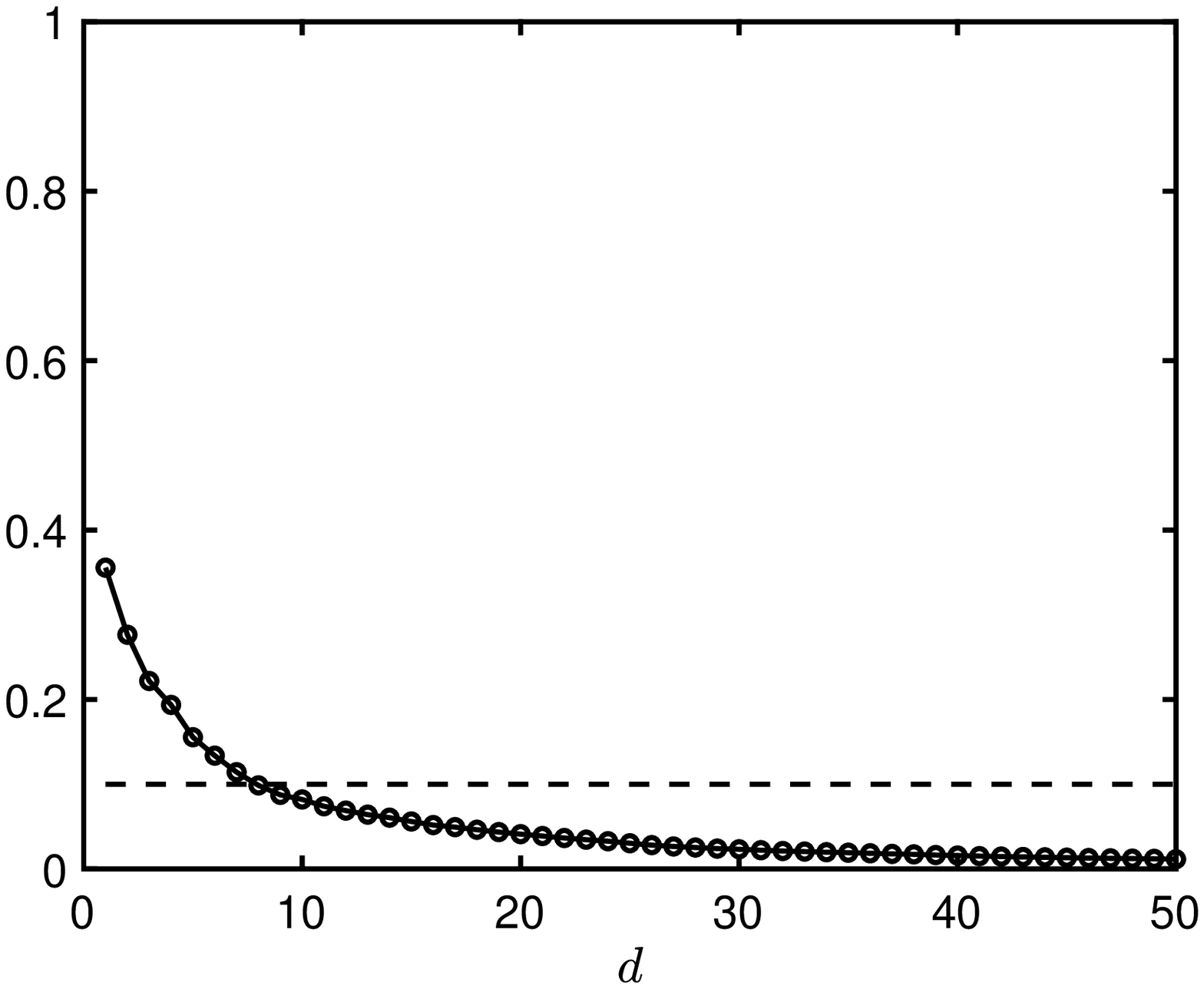}}
    \caption{Relative error $\Vert \tilde{\tensor C}-\tensor C\Vert_F/\Vert \tensor C \Vert_F$ with respect to different $d$ for different datasets. The dash line corresponds to 10\% relative error. (a)A1. (b)B1. (c)C1.}
    \label{fig:dim_red}
\end{figure}

Figure~\ref{fig:dim_red} illustrates the relative error $\Vert \tilde{\tensor C}-\tensor C\Vert_F/\Vert \tensor C \Vert_F$ of the training data sets of A1, B1, and C1. In all cases, the relative error is smaller than $10\%$. This indicates that we only need to identify $8$ critical functional groups to characterize the data sets B1 and C1 when predicting solvation free energy, which implies that these data sets have very good low dimension structure. We also notice that for data set A1, we need $d=25$. This is consistent with the previous BCI analysis. As in data set A1, there are more types of elements (nodes). When we try to identify the critical functional groups/molecular fragments of the data set with model reduction approach, the effect of nodes (elements) on the number of critical groups is stronger than the topology of a molecule.
Even though we do not have a strategy to identify specific functional groups at the moment, the data analysis above shows potential for achieving effective dimension reduction for molecules on solvation free energy prediction. We also note that, because the distance matrix of points in $\mathbb{R}^d$ is invariant under drift or rotation, identifying the map between basis in $\mathbb{R}^d$ and the critical functional groups requires comprehensive investigation and delicate design, which will be a target of our future work. In this work, we only show this potential via providing an abstract proof of concept in mathematics. This method is also valuable for predicting other properties.

%\section*{Discussion}
\section*{Conclusion}
In this work, we introduced a GPR model for solvation free energy prediction. The proposed GPR model used a marginalized graph kernel. A new similarity metric between molecules is defined in the marginalized graph kernel by both molecular topology and geometry. Therefore, the kernel can naturally adapt to molecules containing topological diversity and various types of elements.
We benchmarked the performance of the GPR model on solvation free energy prediction across three data sets. To investigate the effect of different components in solvation free energy calculation as the effect solvent and contribution of conformational entropy, three solvation free energy data sets of our DFT calculations with implicit water model, a subset of QM9 database of MD simulation with implicit water model and a subset of experiment data in Freesolv database were built. We demonstrated that by tuning the hyperparameters, the uncertainty that was generated by explicit solvent and/or conformation change does not affect the accuracy of our GPR model. And we found that our GPR model with the marginalized graph kernel can predict solvation free energy at chemical accuracy (\textless $1$ kcal/mol) for the subsets of QM9 database and Freesolv database while using significantly small training data set (3\% of QM9 database). Wu et al. have noticed that generally, the performance of graph-based model is better than other methods, but is not robust enough on complex tasks under data scarcity. We also identified the same issue for our ML model on data set A1. The complexity of these data sets were further analyzed by model reduction method. We also found that the Bertz complexity index can be used to describe the data scarcity in high dimensional space to some extent. Finally, we showed a new method to evaluate the similarity between molecule in new test set and training set as well as the property prediction, which based on the distance between molecular graphs. This method provides a possible way on which to build a minimum training set to improve prediction for certain test sets. 
The current results show good performance of our GP model with graph kernel. Next step we will combine the current ML model with more descriptors to provide effective guidance for the inverse molecule design of organic molecules in a redox flow battery.
\section{Data and Software Availability}
All the training and test sets in this work are available with the paper (see the SI files). The GP model is build by scikit-learn library version 0.20.3 (https://scikit-learn.org/). The graph kernel is implemented by our GPU-accelerated python library Graphdot version 0.3.2 (https://github.com/yhtang/GraphDot). Additional data or code would be available upon reasonable request.

\begin{acknowledgement}
This work was supported by the Energy Storage Materials Initiative (ESMI), which is a  Laboratory Directed Research and Development Project at Pacific Northwest National Laboratory (PNNL). PNNL is a multiprogram national laboratory operated for the U.S. Department of Energy (DOE) by Battelle Memorial Institute under Contract no. DE- AC05-76RL01830
\end{acknowledgement}
%\section*{Author contributions}
%P.G., X.Y., Y. T., and M. Z. designed and implemented the algorithms. P. G. and X. Y. trained the models and analysed the results. A. A. did DFT calculations of solvation energy. A. H. aided in analyzing data. V. M and W. W. supervised and directed the project. 
\begin{suppinfo}
All the data sets for the machine learning model during this study are included in the Supplementary Information Files.
\end{suppinfo}
%All other relevant source data are available from the corresponding author upon reasonable request.

%\input{support}

\bibliography{ref}
\end{document}